\documentclass[a4paper,fleqn,usenatbib]{mnras}

\usepackage{newtxtext,newtxmath}

\usepackage[T1]{fontenc}
\usepackage{ae,aecompl}

\usepackage{graphicx}	
\usepackage{amsmath}	
\usepackage{amssymb}	

\usepackage{comment}
\usepackage{enumerate}

\usepackage{rotating}
\usepackage{array}
\usepackage{multirow}
\usepackage{longtable}
\usepackage{threeparttable}
\usepackage{float}
\usepackage{color}

\def\app#1#2{%
  \mathrel{%
    \setbox0=\hbox{$#1\sim$}%
    \setbox2=\hbox{%
     \rlap{\hbox{$#1\propto$}}%
      \lower1.1\ht0\box0%
    }%
    \raise0.25\ht2\box2%
  }%
}
\def\approxprop{\mathpalette\app\relax}

\title[Empirical Limits of the Baryon-Cycle Paradigm]{Galaxy And Mass Assembly (GAMA): Gas Fuelling of
Spiral Galaxies in the Local Universe II. -- Direct Measurement of the Dependencies on Redshift and Host Halo Mass of Stellar Mass Growth in Central Disk Galaxies}

\author[M. W. Grootes et al.]{
M.~W.~Grootes$^{1 \dagger}$\thanks{Corresponding authors:\newline mgrootes@cosmos.esa.int, richard.tuffs@mpi-hd.mpg.de},
A.~Dvornik$^{2}$,
R.~J.~Laureijs$^{1}$,
R.~J.~Tuffs$^{3 {\color{blue}\star}}$,
C.~C.~Popescu$^{4,5}$,
\newauthor A.~S.~G.~Robotham$^{6,7}$,
J.~Liske$^{8}$,
M.~J.~I.~Brown$^{9}$,
B.~W.~Holwerda$^{10}$,
L.~Wang$^{11,12}$\\
$^{1}$ESA/ESTEC SCI-S, Keplerlaan 1, 2201 AZ, Noordwijk, The Netherlands\\
$^{\dagger}$ESA Fellow\\ 
$^{2}$Leiden Observatory, University of Leiden, Niels Bohrweg 2, 2333 CA Leiden, The Netherlands\\
$^{3}$Max-Planck-Institut f\"ur Kernphysik, Saupfercheckweg 1, 69117 Heidelberg, Germany\\
$^{4}$Jeremiah Horrocks Institute, University of Central Lancashire, Preston PR1 2HE, UK\\
$^{5}$The Astronomical Institute of the Romanian Academy, Str, Cutitul de Argint 5, Bucharest, Romania\\
$^{6}$University of Western Australia, Stirling Highway Crawley, WA 6009, Australia\\
$^{7}$International Centre for Radio Astronomy Research (ICRAR), University of Western Australia, Stirling Highway Crawley, WA 6009, Australia\\
$^{8}$Hamburger Sternwarte, Universit\"at Hamburg, Gojensbergweg 112, 21029 Hamburg,  Germany\\
$^{9}$School of Physics and Astronomy, Monash University, Clayton, Victoria 3800, Australia\\
$^{10}$University of Louisville, Department of Physics and Astronomy, 102 Natural Science Building, Louisville KY 40292,USA\\
$^{11}$SRON Netherlands Institute for Space Research, Landleven 12, 9747 AD, Groningen, The Netherlands\\
$^{12}$Kapteyn Astronomical Institute, University of Groningen, Postbus 800, 9700 AV, Groningen, The Netherlands    
}

\date{Accepted XXX. Received YYY; in original form ZZZ}

\pubyear{2015}

\begin{document}
\label{firstpage}
\pagerange{\pageref{firstpage}--\pageref{lastpage}}
\maketitle

\begin{abstract}
We present a detailed analysis of the specific star formation rate -- stellar mass ($\mathrm{sSFR}-M_*$) of $z\le 0.13$ disk central galaxies using a morphologically selected mass-complete sample ($M_* \ge 10^{9.5} M_{\odot}$). Considering samples of grouped and ungrouped galaxies, we find the $\mathrm{sSFR}-M_*$ relations of disk-dominated central galaxies to have no detectable dependence on host dark-matter halo (DMH) mass, even where weak-lensing measurements indicate a difference in halo mass of a factor $\gtrsim5$. We further detect a gradual evolution of the $\mathrm{sSFR}-M_*$ relation of non-grouped (field) central disk galaxies with redshift, even over a $\Delta z \approx 0.04$ ($\approx5\cdot10^{8}\mathrm{yr}$) interval, while the scatter remains constant. This evolution is consistent with extrapolation of the "main-sequence-of-star-forming-galaxies" from previous literature that uses larger redshift baselines and coarser sampling.\newline
Taken together, our results present new constraints on the paradigm under which the SFR of galaxies is determined by a self-regulated balance between gas inflows and outflows, and consumption of gas by star-formation in disks, with the inflow being determined by the product of the cosmological accretion rate and a fuelling-efficiency -- $\dot{M}_{\mathrm{b,halo}}\zeta$.  In particular, maintaining the paradigm requires $\dot{M}_{\mathrm{b,halo}}\zeta$ to be independent of the mass $M_{\mathrm{halo}}$ of the host DMH. Furthermore, it requires the fuelling-efficiency $\zeta$ to have a strong redshift dependence ($\propto (1+z)^{2.7}$ for $M_*=10^{10.3} M_{\odot}$ over $z=0 - 0.13$), even though no morphological transformation to spheroids can be invoked to explain this in our disk-dominated sample. The physical mechanisms capable of giving rise to such dependencies of $\zeta$ on $M_{\mathrm{halo}}$ and $z$ for disks are unclear.  
\end{abstract}




\section{Introduction}\label{introduction}
Over the past decade, a wide range of observational work has established the existence of a tight relation between the star formation rate (SFR, or $\Phi_*$) and the stellar mass ($M_*$) of star forming galaxies, with this relation having been in place at least as early in the history of the Universe as $z\sim2.5$ and maybe even as early as $z\sim6$ \citep{NOESKE2007,ELBAZ2007,WUYTS2011,WHITAKER2012,SPEAGLE2014}. This relation -- widely referred to as the 'Main Sequence of Star Forming Galaxies' (MS) -- takes the form of a power-law with normalization and slope evolving with redshift $z$
, while the scatter \footnote{in the sense of the $1-\sigma$ dispersion of galaxies around the MS relation} remains roughly constant at $\sim0.3\,$dex. Notably, it has also been demonstrated that the MS is preferentially populated by disk dominated galaxies, and has been so since at least $z\sim2$ \citep{WUYTS2011}. Accordingly, the majority of stars that have formed in the Universe since at least the peak of the cosmic star formation history at $z\approx1.9$ \citep{MADAUDICKINSON2014} have condensed out of cold gas distributed over the disks of spiral galaxies. It may thus be argued that the physically more fundamental relation underlying the MS relation is the SFR--$M_*$ relation for disk galaxies; connecting their ability to sustain extended star formation to their rotationally supported kinematic structure \citep{DRIVER2006,ABRAMSON2014,GROOTES2014,GROOTES2017}. Given observational evidence implying that the gas required to sustain this process is supplied via continuous accretion from the inter-galactic medium (IGM) \citep[e.g.][]{LHUILLIER2012, ROBOTHAM2014}, the processes regulating this `gas-fuelling' are central to our understanding of galaxy formation and evolution.\newline

Under the present paradigm of structure formation, galaxies initially form and evolve as disk galaxies at the centre of dark matter halos (DMH) \citep[e.g.][]{REES1977,WHITE1978,FALL1980,WHITE1991,MO1998}. Their subsequent evolution is determined by the on-going formation of stars from the inter-stellar medium (ISM) following the Schmidt-Kennicutt relation \citep{SCHMIDT1959,KENNICUTT1998a}. The availability of ISM, in turn, is expected to be determined by a balance between flows of gas into the galaxy, and removal and consumption of the ISM by outflows and star formation, respectively\citep[e.g.][]{RASERATEYSSIER2006,FINLATORDAVE2008,BOUCHE2010,DUTTON2010,DAVE2012,LILLY2013}, i.e. a baryon cycle.

In this picture the star formation rate of a galaxy is set by the interplay and the evolving balance of (i) the rate at which the gas flows into the ISM, (ii) feedback from energetic processes in the galaxy (including star formation) driving outflows of ISM from the galaxy and disrupting flows of incoming gas \citep[e.g.][and references therein]{FAUCHER-GIGUERE2011,HOPKINS2013}, and (iii) the efficiency with which ISM is converted into stars.

Of these three, the latter two are assumed to depend largely on galaxy-specific processes and properties (e.g. star-formation and SNe feedback, galaxy mass, and metallicity), while the inflow rate is (predominantly) expected to depend both on the cosmological epoch as well as on the mass of the galaxy's host dark matter halo (DMH). While the cosmological epoch influences the prevalence of gas via the cosmological accretion rate of DM and baryons from the inter-galactic medium (IGM) onto DMHs \citep[e.g.][]{GENEL2008,MCBRIDE2009}, the mass of the dark matter halo sets the (mix of) accretion mode(s), i.e. `cold mode' accretion from filamentary flows \citep[e.g.][]{KERES2005,DEKEL2009,BROOKS2009,KERES2009,PICHON2011,NELSON2013} and/or `hot mode' accretion from a hot/warm virialized intra-halo medium\footnote{Of course, to be accreted into the ISM of a galaxy the gas being accreted must be cold in the sense that its thermal velocity must be lower than the escape velocity of the ISM. `Cold mode' and `hot mode' refers to the temperature history of the gas, with cold mode accretion consisting of gas that has never been shock heated to temperatures of comparable to the virial temperature, but instead has always remained in a cold, dense state, while hot mode refers to gas that has been shocked and heated and has subsequently cooled.} (IHM;  \citealp[e.g.][]{KERES2005,DEKEL2006,VANDEVOORT2011a,DEKEL2013}). Theory predicts a transition between the two modes at DMH masses of $\sim10^{12} M_{\odot}$ and a further decline of the propensity of gas to cool and be accreted in the hot mode with increasing DMH mass \citep{BIRNBOIM2003,KERES2005,DEKEL2006,BENSON2011,VANDEVOORT2011}.

Accordingly, one expects a gradually evolving, inflow-driven, self-regulated, balance of ISM content and star-formation, at least for central disk galaxies; for satellite galaxies inflows are predicted to be curtailed by the stripping of cold and cooling gas resulting from the motion of the galaxy with respect to the host DMH \citep{GUNN1972,ABADI1999,HESTER2006,BAHE2015,LARSON1980,KIMM2009}. Indeed, implementations of the baryon cycle, both in a sophisticated emergent manner in the form of hydrodynamic simulations \citep[e.g][]{KERES2005,SCHAYE2010,CRAIN2009,HOPKINS2014,SCHAYE2015} and in semi-analytic models of galaxy evolution \citep{COLE2000,LACEY2008,LAGOS2011a, CROTON2006,GUO2011,HENRIQUES2015}, as well as in a simplified analytic form \citep{FINLATORDAVE2008,BOUCHE2010,DAVE2012,LILLY2013,PENG2014}, successfully recover the qualitative behaviour of the observed MS relation, lending credence to the baryon cycle and self-regulated feedback in disk galaxies as the underlying driver for the evolution of galaxies on the MS.\newline     

However, in a recent analysis focussing on isolating and empirically constraining the process of gas-fuelling in disk galaxies in a range of environments, we have shown that the gas-fuelling of these objects is largely independent of the satellite/central dichotomy. In addition this analysis has shown that, even for central galaxies, the environment (group vs. field a proxy for DMH mass) seems to have a negligible impact on their gas-fuelling and star formation (\citealp[][henceforth Paper I]{GROOTES2017}\defcitealias{GROOTES2017}{Paper I}). Both findings are contrary to the theoretical expectations outlined above, and indicate that our understanding of the processes governing gas-fuelling and determining the baryon cycle remains incomplete.\newline

The aim of this paper is therefore to empirically test and constrain the elements of the baryon cycle (of central disk galaxies) and the resulting picture of a an inflow-driven self-regulated star formation rate. In particular, we focus on the evolution of the specific SFR--$M_*$ ($\psi_*$--$M_*$) relation of central disk galaxies over short redshift intervals ($\Delta z \approx 0.04$) in the local Universe, as well as on the dependence on host DMH mass at fixed redshift and stellar mass. 

Under the reasonable assumption that the physical processes regulating star formation in galaxies remain constant, in nature and efficiency, in the local Universe (for $0 \le  z \le 0.13$), the former will enable us to identify variations in galaxy SFR as a result of a gradually evolving inflow/supply of gas and to isolate these from potential variations in the galaxy specific processes such as star formation/feedback. Conversely the latter will enable us to directly constrain the postulated DMH mass dependence and to test in detail to which degree the baryon-cycle and gas-fuelling of central galaxies is impacted by the group environment thus following up our unexpected result from \citetalias{GROOTES2017}.\newline

We make use of the samples and methodology defined and described in detail in \citetalias{GROOTES2017}, and briefly recapitulate the data products and samples used in the analysis in section~\ref{dataandsamples}. In section~\ref{results_evol}  we present our results on the redshift evolution of the  $\psi*$--$M*$ relation for central disk/spiral galaxies, followed by the results of our investigation of the DMH mass dependence (section~\ref{results_env}). We discuss the direct implications of our results for the gas-fuelling and the baryon cycle of central disk galaxies in section~\ref{IMPGF}  and discuss their broader implications in section~\ref{discussion}. Finally a summary and conclusions are presented section~\ref{summary}.\newline
 
Throughout the paper, except where stated otherwise, we make use of magnitudes on the AB scale \citep{OKEGUNN1983} and an $\Omega_M = 0.3$, $\Omega_{\lambda} = 0.7$, $H_0 = 70\,\mathrm{km}\mathrm{s}^{-1}\mathrm{Mpc}^{-1}$ ($h=0.7$)  cosmology. \newline 

\section{Data \& Samples}\label{dataandsamples}
As in \citetalias{GROOTES2017}, the Galaxy And Mass Assembly survey \citep[GAMA][]{DRIVER2011,LISKE2015} forms the basis for our analysis. In addition to the combined spectroscopic and multi-wavelength broadband imaging data from the far ultra-violet (FUV) to the far infra-red (FIR), GAMA also provides a wide range of ancillary data products, including, but not limited to, emission line measurements \citep{HOPKINS2013}, aperture matched \citep{HILL2011} and single S\'ersic profile photometry in the optical--NIR \citep[][with associated structural parameters]{KELVIN2012}, UV photometry (\citealp{LISKE2015}, Andrae et al., in prep.), stellar mass estimates \citep{TAYLOR2011}, and a highly complete friends-of friends group catalogue \citep{ROBOTHAM2011b}.\newline

In \citetalias{GROOTES2017} we used these data products to define 
volume limited, morphologically selected samples of local universe ($z\le0.13$) disk galaxies, including samples of field and group central disk galaxies. For these, we use GAMA's NUV photometry in combination with a novel radiation-transfer-model-based attenuation correction technique \citet{POPESCU2011,GROOTES2013}, to derive precise and accurate intrinsic total star formation rates (SFR) as a tracer of gas content.\newline

We refer the reader to \citet{DRIVER2011,LISKE2015,DRIVER2016} and references therein, as well as to the references provided above, for details of the GAMA survey and the individual data products. Furthermore, we refer the reader to \citetalias{GROOTES2017}, for a detailed synopsis of the derived properties used in this analysis, including in particular stellar mass and star-formation rate, as well as for a full description of the sample selection. In the following, however, we briefly outline the most salient details.\newline

\subsection{Data \& Derived Physical Properties}\label{dataandderivedprop}
Our analysis uses the first 3 years of data of the GAMA survey - frozen and referred to as GAMA I - consisting of the three equatorial fields to a homogeneous depth of $r_{\mathrm{AB}} \le 19.4\,$mag\footnote{the $r$-band magnitude limit for the GAMA survey is defined as the SDSS Petrosian foreground extinction corrected $r$-band magnitude}. We make use of GAMA's quantitative spectroscopy as well as of the UV/optical (NUV,$u$,$g$,$r$,$i$,$z$) broadband photometry.\newline

\subsubsection{Quantitative Spectroscopy and the Galaxy Group catalogue}\label{specandg3c} 
GAMA provides spectroscopy and derived quantities, including emission line fluxes, for for $>98$\% of $r<19.4$ galaxies in the survey area. The spectroscopy enables (i) identification and removal of disk galaxies hosting AGNs using the \citet{KEWLEY2001} BPT criterion, and (ii) 
the construction of the GAMA galaxy group catalogue (G$^3$C, \citealp{ROBOTHAM2011b}). Uniquely, as a result of GAMA's high spectroscopic completeness even on small angular scales, the G$^3$C  reliably samples the DMH mass function down to low mass ($M_{\mathrm{dyn}} < 10^{12}$), low multiplicity ($N < 5$), galaxy groups. This catalogue also provides an estimate of the parent halo mass based on a group's total luminosity, which has been cross calibrated using weak-lensing measurements of the group halo mass and the GAMA survey mocks \citep{MERSON2013,HAN2015}. \newline

\subsubsection{Optical Photometry, Stellar Masses, and Weak-Lensing}\label{optandsm}
Homogenized optical photometry -- $u,g,r,i,z,$, based on imaging by the Kilo Degree Survey (hereafter referred to as KiDS; \citealp{KUIJKEN2015,HILDEBRANDT2017,DEJONG2017} ) and archival imaging data of SDSS -- is available for the entire GAMA I footprint. This has enabled the construction of a catalogue of aperture matched photometry \citet{HILL2011,DRIVER2016,WRIGHT2016} as well as of a catalogue of single S\'ersic photometry and structural measurements \citep{KELVIN2012}, providing measurements of effective radii, integrated luminosity, and S\'ersic index in each band for the vast majority of GAMA sources. Foreground extinction corrections in all optical bands have been calculated following \citet{SCHLEGEL1998} and k-corrections to $z = 0$ have been calculated using \verb kcorrrect_v4.2  \citep{BLANTON2007}.\newline

The optical photometry also represents the basis of GAMA's stellar mass measurements following \citep{TAYLOR2011}. These estimates make use of a \citet{CHABRIER2003} IMF and the \citet{BRUZUAL2003} stellar population library. \citet{TAYLOR2011} determine the formal random uncertainties on the derived stellar masses to be $\sim0.1-0.15\,$dex on average, and the precision of the determined mass-to-light ratios to be better than $0.1\,$dex.\newline

Finally, the overlap of GAMA with the KiDS surveys allows us to perform a stacked weak lensing analysis of the DMHs hosting the galaxies from our samples, thus extracting mass estimates. This is discussed in greater detail in section~\ref{avDMHmass} and appendix~\ref{APPEND_WL}. The KiDS data used for this purpose are processed by THELI \citep{ERBEN2013} and Astro-WISE \citep{BEGEMAN2013,DEJONG2015}. Shears are measured using lensfit \citep{MILLER2013}, and photometric redshifts are obtained from PSF-matched photometry and calibrated using external overlapping spectroscopic surveys \citep[see][]{HILDEBRANDT2016}.\newline

\subsubsection{UV Photometry and SFR}\label{uvandsfr}
The majority of the GAMA I footprint has been observed in the NUV by GALEX to a depth of $\sim23\,\mathrm{mag}$ by the MIS \citep{MARTIN2005,MORRISSEY2007} survey and by a dedicated guest investigator 
program \textit{GALEX-GAMA} providing largely homogeneous coverage. This forms the basis for GAMA's NUV photometry. Details of the GAMA UV photometry are provided in \citet{LISKE2015}, Andrae et al. (in prep.), and on the GALEX-GAMA website\footnote{www.mpi-hd.mpg.de/galex-gama/}, and a detailed synopsis is provided in \citetalias{GROOTES2017}. Foreground extinction corrections and k-corrections having been applied as in the optical bands.\newline 

As detailed in \citetalias{GROOTES2017}, the integrated NUV emission from a spiral/disk galaxy provides a star-formation rate tracer  which is sensitive to the total SFR of the galaxy on timescales of $\lesssim10^8\,$yr (see e.g. Fig.~1 of \citetalias{GROOTES2017}), while remaining robust against stochastic fluctuations, unlike H$\alpha$ based tracers. Thus, the timescale probed by the NUV is short compared to the timespan corresponding to a redshift baseline of $\Delta z\approx0.04$ (in the range $z=0-0.13$), making it well suited to investigate the evolution of the $\psi_*$--$M_*$ relation. In this paper we have adopted the calibration between NUV luminosity and SFR as given in \citet{HAO2011},
scaled from a \citet{KROUPA2001} IMF to a \citet{CHABRIER2003} IMF as in \citet{SPEAGLE2014}\footnote{This choice is different from that adopted in \citetalias{GROOTES2017}, and is motivated solely by reasons of inter-comparability, and only significantly impacts the normalization of the $\psi_*$--$M_*$.} .\newline

To correct for the attenuation of stellar emission by dust in the galaxy, which is particularly sever at short (UV) wavelengths \citep[e.g.][]{TUFFS2004}, we employ the method of \citet{GROOTES2013}. This method makes use of the radiation transfer model of \citet{POPESCU2011} and supplies attenuation corrections on an object-by-object basis for spiral galaxies, taking into account the orientation of the galaxy in question and estimating the disk opacity from the stellar mass surface density. A comparison of the method and its performance with a range of other widely used SFR indicators can be found in \citet{DAVIES2016}.\newline

\subsection{Samples of Central Disk Galaxies}\label{samples}
We make use of the sample of field central disk galaxies and the sample of group central disk galaxies as provided in \citetalias{GROOTES2017}. The samples are constructed by morphologically selecting disk galaxies from the full GAMA sample using the method described in \citet{GROOTES2014}, resulting in samples which are unbiased in their star formation properties and are jointly optimized for purity and completeness and. We impose a redshift limit of $z=0.13$, resulting in a mass-complete sample for galaxies with $M_* \ge 10^{9.5}$, and deselect galaxies hosting an AGN based on their position in the BPT diagram. In our analysis we only make use of the mass complete sample, but do, in some cases, include galaxies below this mass on plots to indicate trends in the population.\newline

\subsubsection{Field Central Disks (\textsc{FCS})}\label{fieldcentral}
\begin{figure}
\includegraphics[width=\columnwidth]{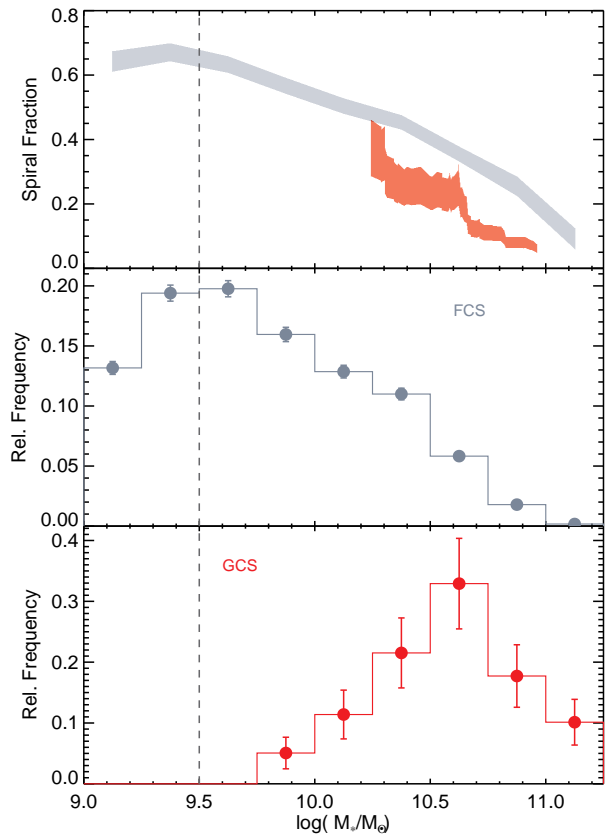}
\caption{Disk fraction as a function of stellar mass $M_*$ (top) and stellar mass functions of the \textsc{FCS} (middle; gray) and \textsc{GCS} samples (bottom; red). The disk fraction of the \textsc{FCS} sample is determined in bins of $0.25\,$dex in $M_*$ with the width of the shaded area indicating the bootstrapped uncertainty in the median. The disk fraction for the \textsc{GCS} sample is determined in a sliding tophat bin with bounds defined such as to encompass 25 \textsc{GCS} galaxies. As for the \textsc{FCS} sample, the width of the shaded area denotes the uncertainty in the median. The mass functions are shown in bins of $0.25\,$ dex in stellar mass for both samples, with Poisson errors on the relative frequencies. Although the \textsc{FCS} and \textsc{GCS} samples display a mutual range of stellar mass above $M_* = 10^{9.8} M_{\odot}$, the mass functions very different. For the \textsc{GCS} sample we find a median stellar mass of $M_* = 10^{10.6} M_{\odot}$. The disk fraction amongst the \textsc{GCS} sample is lower than that of the \textsc{FCS} sample by $\approx 0.2 -0.3$ at all $M_*$.}
\label{fig_FGCS_spfsm}
\end{figure}

From  the parent sample of disk galaxies, a sample of field central disk galaxies is selected as those which are not associated with any other galaxy to the limiting depth of the survey by the Friends-of-Friends group finding algorithm \citep{ROBOTHAM2011b}. As such, these galaxies likely represent the dominant central galaxy of their DMH, with any satellite being at least less massive than $M_* = 10^{9.5} M_{\odot}$ and likely even less massive over most of the redshift range. In the following we will refer to this sample, encompassing 3508 
galaxies, as the \textsc{FCS} sample\footnote{In \citetalias{GROOTES2017} we referred to this sample as the \textsl{\textsc{fieldgalaxy}} sample.}. Fig.~\ref{fig_FGCS_spfsm} shows the fraction\footnote{The disk fraction is calculated relative to the super-sample of galaxies which meet all the requirements of the \textsc{FCS} sample, save for the morphological requirements and the AGN de-selection. The impact of the latter criterion is negligible ($<1$\%).} of disk galaxies in the field as a function of stellar mass, as well as the stellar mass distribution of the \textsc{FCS} sample.\newline

\subsubsection{Group Central Disks (\textsc{GCS})}\label{groupcentrals} 
In selecting a sample of group central disk galaxies, we proceed by selecting those galaxy groups from the the G$^3$C which contain at least three member galaxies with $M_* \ge 10^{9.5} M_{\odot}$ (regardless of the galaxies' morphology) and again impose the redshift limit of $z=0.13$. This results in a volume limited sample of galaxy groups. From these, we then select those galaxies which are the central galaxy of the group, are a member of the parent sample of disk galaxies, and have no neighbouring galaxy within $50\,\mathrm{kpc}\,h^{-1}$ and $1000\,\mathrm{km}\mathrm{s}^{-1}$. This latter criterion is imposed to ensure that the SF activity of the galaxy is unlikely to be impacted by galaxy-galaxy interactions, which are known to influence the SFR and SF efficiency of galaxies (\citealp[e.g.][]{BARTON2000,ROBOTHAM2013,ROBOTHAM2014,DAVIES2015,ALATALO2015,BITSAKIS2016} ). In the following we will refer to this sample of 79 largely isolated  group central disk galaxies as the \textsc{GCS} sample. For reference, we show the disk fraction\footnote{The central disk fraction is calculated relative to the total population of group centrals of the volume limited group sample, i.e including AGN hosts and galaxies with a neighbour within the separation criteria. The de-selection of these galaxies increases the down-selection to the \textsc{GCS} sample by $<10$\%.}. The stellar mass distribution is clearly skewed towards more massive galaxies for the \textsc{GCS} sample than for the \textsc{FCS} sample, with the distribution being peaked at the median value of $M_* = 10^{10.6} M_{\odot}$.\newline 

\section{Redshift evolution of the $\psi_*$--$M_*$ relation of field central disk galaxies}\label{results_evol}
The $\psi_*$--$M_*$ relation for field central\footnote{As discussed in \citetalias{GROOTES2017} only $\sim20$\% of disk-dominated galaxies at a given $M_*$ are are not field central galaxies (the majority of these are satellite galaxies), resulting in the $\psi_*$--$M_*$ of disk galaxies being dominated by field central galaxies. Furthermore, as also shown in \citetalias{GROOTES2017}, satellite disk galaxies follow a relation similar to that of their central counterparts, albeit with a larger scatter.} disk galaxies likely underlies the so-called `main sequence of star-forming galaxies' \citep{NOESKE2007,WUYTS2011,WHITAKER2012,SPEAGLE2014}, a cornerstone empirical result of galaxy evolution studies of the past decade. Although a recent meta-analysis by \citet{SPEAGLE2014} has calibrated a smooth parameterization of the evolution of the MS relation over the redshift range $z=0.25 - 4$, a probe of its actual smoothness, i.e. the continuity of its evolution over (very) short redshift intervals (and at very low $z$) -- desirable in terms of constraints on the contribution and importance of different physical processes to the relation and its evolution -- remains lacking.\newline

\begin{figure}
\includegraphics[width=\columnwidth]{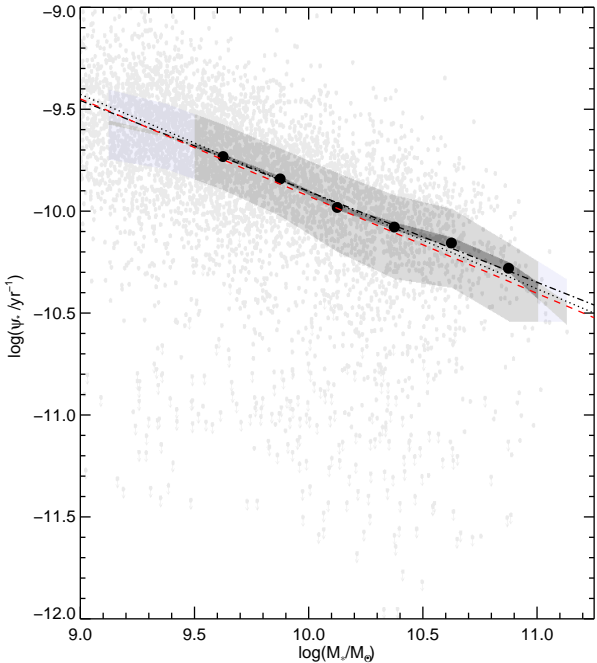}
\caption{Specific star formation rate $\psi_*$ as a function of stellar mass $M_*$ for the \textsc{FCS} sample. Values for individual sample galaxies are shown as gray circles, with downward arrows indicating those galaxies for which the derived value of $\psi_*$ is an upper limit (at $2.5-\sigma$; as discussed in detail in \citetalias{GROOTES2017} the depth of the GALEX-GAMA UV data is such, that the second quartile and the median are defined by detections.). The median $\psi_*$--$M_*$ relation in bins of $0.25\,$dex in $M_*$ is shown by the transparent shaded regions, with the width of the darker narrower region indicating the boot-strapped uncertainty in the median, and the width of the  wider, lighter region indication the inter-quartile range. Given the mass limit of $M_* \ge 10^{9.5}$ to which the \textsc{FCS} sample is volume complete, and the low number of sources with $M_* \ge 10^{11} M_{\odot}$, we have fit a single power law $\mathrm{log}(\psi_*) = A + \gamma(\mathrm{log}(M_*) -10)$ to the binwise median values denoted by the black filled circles, denoted by the black dash-dotted line. For comparison, the empirical parameterization of the main sequence of star-forming galaxies presented by \citet{SPEAGLE2014}, extrapolated to the median redshift of the \textsc{FCS} sample of $z=0.1$, is shown as a red dashed line, while a black dotted line shows the result of fitting power-law with the slope fixed to the expectation value for the main sequence.}    
\label{fig_FCS_ssfrsm}
\end{figure}

Fig.~\ref{fig_FCS_ssfrsm} shows $\psi_*$ as a function of $M_*$ for the \textsc{FCS} sample, with the median relation overlaid. As demonstrated by the figure, the $\psi_*$--$M_*$ relation for the \textsc{FCS} sample is well described by a single power-law
\begin{equation} 
\mathrm{log}(\psi_*) = A +\gamma (\mathrm{log}(M_*) -10)
\label{eq_pl_fit}
\end{equation} 
over its entire range in stellar mass, with $\gamma =-0.45\pm0.01$ as in \citetalias{GROOTES2017} (see also Table~\ref{tab_PLFITS}). The relation is in good agreement with the low-z extrapolation of the empirical parameterization of the MS provided by \citet{SPEAGLE2014}, shown in red. Both the power-law slope $\gamma$ and the normalization constant $A$ agree with the corresponding values of the parameterization of \citet{SPEAGLE2014} at the mean redshift of the \textsc{FCS} sample ($z=0.1$) within 2-$\sigma$ of their formal  uncertainties. If we also consider the uncertainties on the predicted parameters then the power-law fits are consistent with the predictions for the MS within the 1-$\sigma$ uncertainties of the latter. The details of the fitted power-laws are listed in Table~\ref{tab_PLFITS}, as are the slope and normalization of the parameterization of the MS presented by \citet{SPEAGLE2014}. We do note, however, that within the uncertainties, the fit to the $\psi_*$--$M_*$ has a slightly shallower slope than the parameterization of the MS, with the difference being most noticeable at higher $M_*$. This may result from the inclusion of more bulge-dominated galaxies in the MS sample of \citet{SPEAGLE2014}.\newline

We have established that the median $\psi_*$--$M_*$ relation of central disk galaxies coincides with the parameterization of the MS for our volume-limited sample extending to $z=0.13$ over the full extent in $M_*$ covered. We now investigate the what evolution, if any, occurs in this redshift range. For this purpose we divide the \textsc{FCS} sample into three bins in redshift; $\mathrm{z}_1\,:\,0.03 \le z \le 0.06$, $\mathrm{z}_2\,:\,0.08 \le z \le 0.11$, and $\mathrm{z}_3\,:\,0.12 \le z \le 0.13$. For these three sub-samples we find mean redshifts of $\overline{z_1}=0.05$, $\overline{z_2}=0.095$, and $\overline{z_3} = 0.125$, respectively. \newline

\begin{figure}
\includegraphics[width=\columnwidth]{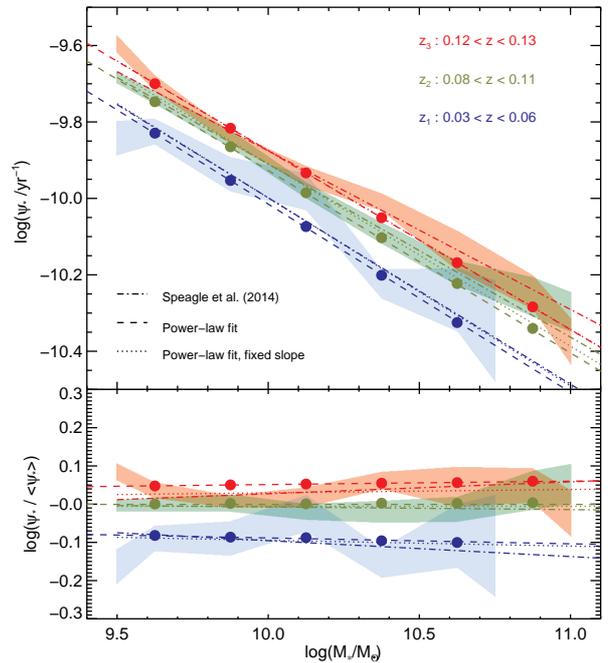}
\caption{Median $\psi_*$--$M_*$ relations for the $\mathrm{z}_1$ (blue, $0.03 \le z \le 0.06$), $\mathrm{z}_2$ (green, $0.08 \le z \le 0.11$), and $\mathrm{z}_3$ (red, $0.12 \le z \le 0.13$) sub-samples of the \textsc{FCS} sample (top). The width of the shaded regions indicate the boot-strapped uncertainty in the median. Single power-law fits of the form $\mathrm{log}(\psi_*) = A + \gamma(\mathrm{log}(M_*) -10)$ to the bin wise median relations are shown as dash-dotted lines in the corresponding color, while dashed lines indicate the extrapolated expectation for the main sequence following \citet{SPEAGLE2014} for the mean redshift of each sub-sample ($\overline{z_1} = 0.05$, $\overline{z_2} = 0.095$, and $\overline{z_3}=0.125$). The binwise expectations following \citet{SPEAGLE2014} are shown as filled circles. For comparison, the results of fitting a single power-law with the slope fixed to that expected for the main sequence is shown as a dotted line. The bottom panel shows the median relations normalized to the single power-law fit to the full \textsc{FCS} sample as shown in Fig.~\ref{fig_FCS_ssfrsm} and listed in Table~\ref{tab_PLFITS} and to the extrapolated expectation for the main sequence at the median redshift of $z=0.1$ for the power-law fits and the main sequence expectations, respectively. Color coding and line-styles are identical to the top panel.}
\label{fig_FCS_ssfrsm_evol}
\end{figure}

The top panel of Fig.~\ref{fig_FCS_ssfrsm_evol} depicts the median $\psi_*$--$M_*$ relation of the three sub-samples binned in stellar mass, as well as the best fit power-law and the predictions of \citet{SPEAGLE2014} (making use of the mean redshift) for each sub-sample. The $\psi_*$--$M_*$ relations for the sub-samples agree well with the extrapolation of the empirical MS parameterization, albeit that small differences in normalization and/or slope are present, which largely averaged out in the full \textsc{FCS} sample. The normalization constants for all sub-samples agree with those expected for the MS within the $1-\sigma$ formal uncertainties of the fits. Similarly, for sub-samples $\mathrm{z}_1$ and $\mathrm{z}_2$ the fitted power-law slopes agree with those of the extrapolated MS within the $1-\sigma$ formal uncertainties of the fit, and only for the $\mathrm{z}_3$ subsample do the slopes differ more, i.e. by $2-\sigma$\footnote{Including the uncertainties in the coefficients of the parameterization, all fitted power-law relations agree with the extrapolated relation of \citet{SPEAGLE2014} within $1-\sigma$ uncertainties both in slope and normalization.}. The full details of the fits are listed in Table~\ref{tab_PLFITS}.\newline

\begin{table*}
	\centering
	\caption{Compilation of power law fits to the $\psi_* - M_*$ relation}
	\label{tab_PLFITS}
	\begin{tabular}{cccccccccc}
 	& & \multicolumn{4}{c}{fit} & \multicolumn{4}{c}{MS expectation} \\
 	Sample & $\overline{z}$ & $A$ & $\gamma$ & $A_{\mathrm{norm}}$ & $\gamma_{\mathrm{norm}}$ & $A$ & $\gamma$ & $A_{\mathrm{norm}}$ & 							$\gamma_{\mathrm{norm}}$  \\
	 \hline
 	\textsc{FCS} & $0.1$ & $-9.90\pm0.01$ & $-0.45\pm0.01$ & - & - & $-9.92$ & $-0.47$ & - & - \\
 	$\mathrm{z}_1$ & $0.05$ & $-10.00\pm0.02$ & $-0.49\pm0.06$ & $-0.1\pm0.02$ & $-0.04\pm0.06$  & $-10.01$ & $-0.49$ & $-0.10$ & $-0.02$ 	\\
	$\mathrm{z}_2$ & $0.095$ & $-9.91\pm0.01$ & $-0.45\pm0.02$ & $-0.01\pm0.01$ & $-0.01\pm0.02$ & $-9.92$ & $-0.47$ & $-0.01$ & $0.00$ 	\\
	$\mathrm{z}_3$ & $0.125$ & $-9.88\pm0.01$ & $-0.42\pm0.02$ & $0.03\pm0.01$ & $0.03\pm0.02$  & $-9.88$ & $-0.46$ & $0.05$ & $0.01$\\ 
	\hline
	\hline
	\end{tabular}
	
	\medskip
	\begin{flushleft}
	Power law fits of the form $\mathrm{log}(\psi_*) = A  + \gamma \cdot \left(\mathrm{log}(M_*) -10\right)$ to the $\psi_* - M_*$ relations 
	for different samples of spiral galaxies. The uncertainties reflect the formal uncertainties of the fit. The columns under \citet{SPEAGLE2014} MS
	expectation provide the extrapolated expectation values for the MS following \citet{SPEAGLE2014}. For the purpose of our comparison we have
	converted the empirical parameterization of the MS from the SFR to the specific SFR $\psi_*$ and have shifted the zeropoint in line with our
	choice for the power-law fits. 
	Their best fit Eq.~28 then takes the form  $\mathrm{log}(\psi_*) = (a -1 + bt)[\mathrm{log}(M_*) -10] - (c + dt)$, where $t$ is
	the age of the Universe and the coefficient values are $a=0.84\pm0.02$, $b=0.026\pm0.003$, $c=6.51\pm0.24$, and $d=0.11\pm0.03$.
	These uncertainties in the fit parameters propagate to uncertainties in the effective slope and normalization predicted by the relation at any
	given time. Here we have chosen to list only the predicted values in the table. For the redshift range of our sample typical uncertainties for the
	normalization and slope are $\delta A \approx 0.43\,$dex and $\delta \gamma \approx 0.04\,$dex. Parameters with subscript `$\mathrm{norm}$' correspond to the power-laws re-fit to the subsamples after normalization to the result obtained for the full \textsc{FCS} sample and the \citet{SPEAGLE2014} expectation for the MS at $z=0.1$, respectively. 
	\end{flushleft}
\end{table*}

As the focus of our interest is on the relative evolution of the relations over the redshift range covered by our sample, in order to facilitate a comparison, we have normalized each sub-sample by the fit to the full \textsc{FCS} sample and have re-fit a power-law. Analogously, we have normalized the extrapolated MS relations at the mean redshift of each sub-sample to the MS relation at $z=0.1$; the results are shown in the bottom panel of Fig.~\ref{fig_FCS_ssfrsm_evol} and the fit parameters $A_{\mathrm{norm}}$ and $\gamma_{\mathrm{norm}}$ are listed in Table~\ref{tab_PLFITS}. We find that all observed differences in normalization are consistent with those predicted for the MS, while also being statistically significant at $>3\,\sigma$, with the exception of $\mathrm{z}_3\rightarrow\ \mathrm{z}_2$, for which the difference in normalization is only $0.04\,$dex ($2-\sigma$). A synopsis of the observed and expected evolution in normalization between the three sub-samples of the \textsc{FCS} sample is provided in Table~\ref{tab_norm}.

In addition to an evolution of the normalization of the MS, the empirical parameterization presented by \citet{SPEAGLE2014} also predicts the slope of the MS to evolve, becoming steeper with decreasing redshift. For the $\mathrm{z}_1$ and $\mathrm{z}_2$ sub-samples the fitted slopes of the power-law relations are consistent with those expected for the MS within the formal uncertainties of the fits (see Table~\ref{tab_PLFITS}, while for the $\mathrm{z}_3$ sub-sample the slope is shallower than the MS expectation by $2-\sigma$, as also visible in the top panel of Fig.~\ref{fig_FCS_ssfrsm_evol}. We note, however, that the evolution in slope is so small, that we can not robustly exclude the scenario of no evolution.\newline

\begin{table}
	\centering
	\caption{Compilation of evolution in normalization and slope for power law fits to the $\psi_* - M_*$ relation}
	\label{tab_norm}
	\begin{tabular}{cccccc}
 	& & \multicolumn{2}{c}{fit} & \multicolumn{2}{c}{MS expectation} \\
 	Sample 1 & Sample 2 & $\Delta$A$_{\mathrm{obs}}$ & $\Delta \gamma_{\mathrm{obs}}$ & $\Delta$A$_{\mathrm{MS}}$ & $\Delta \gamma_{\mathrm{MS}}$ \\
 	\hline
 	$\mathrm{z}_3$ & $\mathrm{z}_1$ & $0.13\pm0.02$ & $0.07\pm0.06$ & $0.15$ & $0.03$ \\
	$\mathrm{z}_3$ & $\mathrm{z}_2$ & $0.04\pm0.02$ & $0.04\pm0.02$ & $0.06$ & $0.01$ \\
	$\mathrm{z}_2$ & $\mathrm{z}_1$ & $0.09\pm0.02$ & $0.03\pm0.06$ & $0.09$ & $0.02$ \\ 
	\hline
	\hline
	\end{tabular}

	\medskip
	\begin{flushleft}
	Observed (obs) and expected (MS) values (for the MS) of evolution in normalization and slope of the $\psi_*$--$M_*$ between redshift sub-samples computed from the normalized power-law fits listed in Table~\ref{tab_PLFITS}. Uncertainties correspond to the formal uncertainties in sum quadrature. As in Table~\ref{tab_PLFITS} we list no uncertainties for the predictions. For the redshift range considered typical values would be $\delta \Delta A \approx 0.6\,$dex and $\delta \Delta \gamma \approx 0.05\,$dex. 	
	\end{flushleft}	
\end{table}

We complement our investigation of the evolution of the median $\psi_*$--$M_*$ by an investigation of the distribution of $\psi_*$ at given $M_*$ . To this end we split each of our redshift sub-samples $\mathrm{z}_i$, into three bins of stellar mass, $\mathrm{M}_1\,:\,9.5 \le \mathrm{log}(M_*/M_{\odot}) <10$, $\mathrm{M}_2\,:\,10 \le \mathrm{log}(M_*/M_{\odot}) <10.5$, and $\mathrm{M}_3\,:\,10.5 \le \mathrm{log}(M_*/M_{\odot}) <11.0$\footnote{the combined mass range $9.5\le \mathrm{log}(M_*/M_{\odot}) <11.0$ encompasses $\gtrsim99$\% of the \textsc{FCS} sample.}, and define the quantities $\Delta \mathrm{log}(\psi_*)$, and $\Delta_{\mathrm{z}_i}\mathrm{log}(\psi_*)$ as
\begin{eqnarray}
\Delta \mathrm{log}(\psi_*) & = & \mathrm{log}(\psi_*) - \mathrm{log}(\left<\psi_{*,\mathrm{FCS}}(M_{*})\right>)\,,\,\mathrm{and}\\
\nonumber \\
\Delta_{\mathrm{z}_i} \mathrm{log}(\psi_*) & = & \mathrm{log}(\psi_*) - \mathrm{log}(\left< \psi_{*,\mathrm{z}_i}(M_{*})\right>)\,,
\end{eqnarray}
respectively, where $\left<\psi_{*,\mathrm{FCS}}(M_{*})\right>$ is the median value of $\psi_*$ at $M_*$ as determined from the full \textsc{FCS} sample and $\left<\psi_{*,\mathrm{z}_i}(M_{*})\right>$ is the corresponding median value as determined from the sub-sample $\mathrm{z}_i$. As such, $\Delta \mathrm{log}(\psi_*)$, and $\Delta_{\mathrm{z}_i}\mathrm{log}(\psi_*)$ quantify the distribution of $\psi_*$ around the median relation(s), thus normalizing out the dependence of $\psi_*$ on $M_*$.\newline

\begin{figure*}
\includegraphics[width=0.9\textwidth]{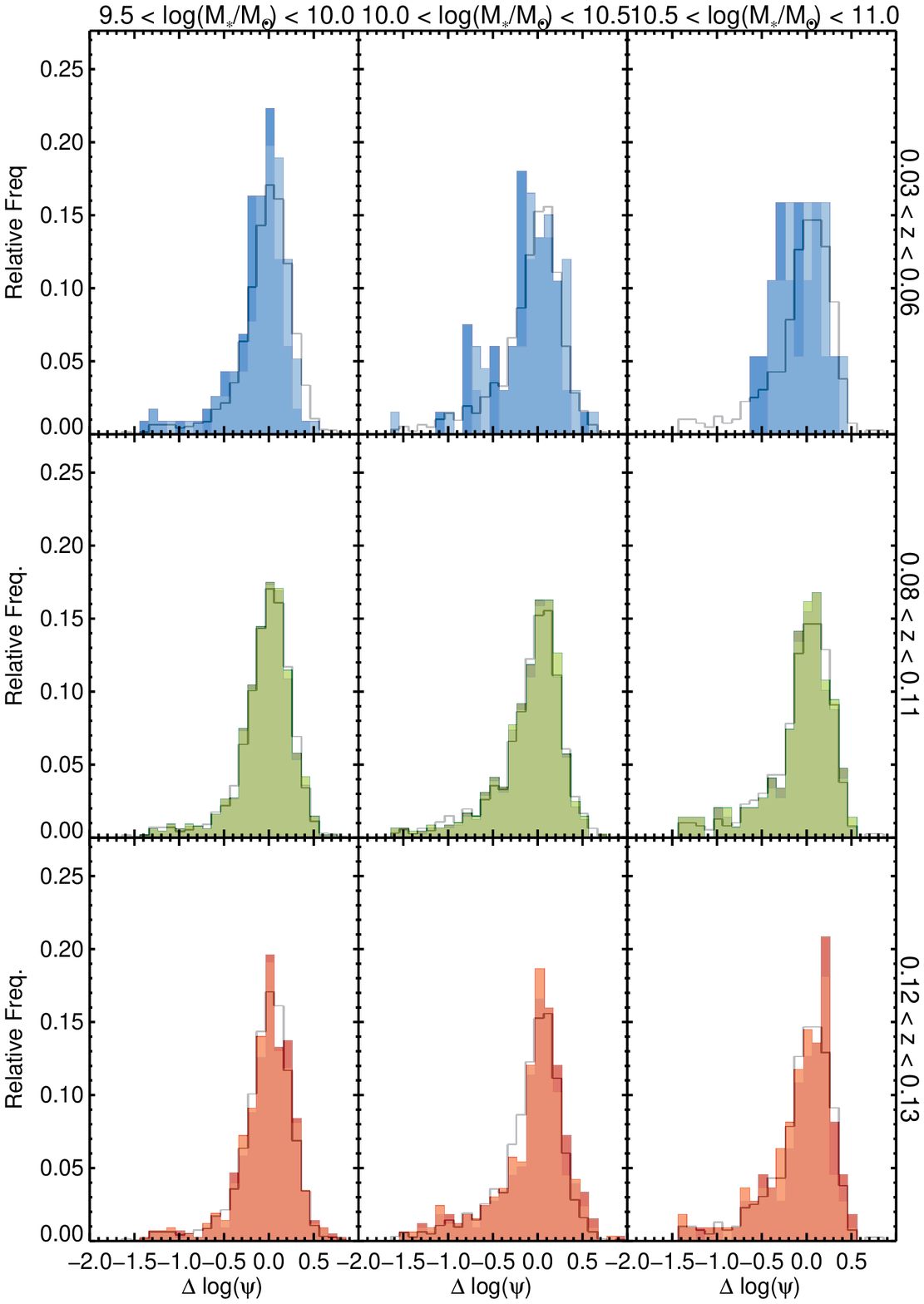}
\caption{Evolution of $\Delta\mathrm{log}(\psi_*)$ for the $\mathrm{z}_1$, $\mathrm{z}_2$, and $\mathrm{z}_3$ sub-samples of the \textsc{FCS} sub-sample (top to bottom) in three bins of stellar mass $M_*$ as indicated (left to right). The mass range encompasses $\gtrsim99$\% of the \textsc{FCS} sample. The distribution of $\Delta\mathrm{log}(\psi_*)$ in each mass bin for the full \textsc{FCS} sample is shown as a gray solid line, while the distributions of $\Delta\mathrm{log}(\psi_*)$ for the redshift subsamples are shown as dark shaded histograms and the distributions of $\Delta_{\mathrm{z}_i}\mathrm{log}(\psi_*)$ are shown as lighter shaded histograms. quantitative statistical tests of the similarity of the distributions within each stellar mass bin are listed in Table~\ref{tab_ADmbin}. }
\label{fig_FCS_dlogpsi_evol}
\end{figure*}

Fig.~\ref{fig_FCS_dlogpsi_evol} shows the distributions of $\Delta \mathrm{log}(\psi_*)$ and $\Delta_{\mathrm{z}_i}\mathrm{log}(\psi_*)$ for each of the redshift sub-samples $\mathrm{z}_i$ in the stellar mass bins $\mathrm{M}_j$. In each stellar mass bin the distributions of $\Delta \mathrm{log}(\psi_*)$ and $\Delta_{\mathrm{z}_i}\mathrm{log}(\psi_*)$ are nigh identical between the $\mathrm{z}_i$ sub-samples, with only a variable offset visible for the distributions in $\Delta \mathrm{log}(\psi_*)$ in line with the previously described evolution of the median. This entails that it is an overall shift of the distribution of $\psi_*$ at fixed $M_*$, rather than a change in shape of the relative distribution of $\psi_*$, which drives the observed redshift evolution of the median $\psi_*$--$M_*$ relation. In a more statistically robust sense, this is corroborated by 3-sample Anderson-Darling tests comparing the distributions of $\Delta \mathrm{log}(\psi_*)$ and $\Delta_{\mathrm{z}_i}\mathrm{log}(\psi_*)$. These find no grounds for rejecting the null hypothesis that the distributions are drawn from the same parent sample, modulo a redshift dependent shift in the normalization (see Table~\ref{tab_ADmbin}). Of course, the lack of difference in a statistical sense does not prove similarity, but these results do reinforce our finding of similarity based on visual inspection of the distributions.\newline

\begin{table*}
	\centering
	\caption{Comparison at fixed $M_*$ of the distributions of $\Delta\mathrm{log}(\psi_*)$ and $\Delta_{\mathrm{z}_i}\mathrm{log}(\psi_*)$ }
	\label{tab_ADmbin}
	\begin{tabular}{cccc}
  	& \multicolumn{3}{c}{$M*$ range}  \\
  	Test & $10^{9.5} M_{\odot} \le M_* < 10^{10} M_{\odot}$ & $10^{10} M_{\odot} \le M_* < 10^{10.5} M_{\odot}$ & $10^{10.5} M_{\odot} \le M_* < 10^{11} M_{\odot}$  \\
  	\hline
  	A & $\lesssim10^{-5}$ & $0.002$  & $0.13$ \\
  	B & $\ge 0.9$ & $0.68$ & $\ge 0.9$ \\
  	C & $0.76$ & $0.79$& $\ge 0.9$ \\
	\hline
	\hline
	\end{tabular}	
	
	\medskip	
	\begin{flushleft}
	$p$-values for the null hypothesis that (A) the distributions of $\Delta\mathrm{log}(\psi_*)$ for each of the sub-samples $\mathrm{z}_i$ in three ranges of $M_*$ are consistent with the samples being drawn from the same parent population, (B) the same as (A) but with each distribution having been shifted by the shift of its median w.r.t. that of the $\Delta\mathrm{log}(\psi_*)$ distribution of the full \textsc{FCS} sample, and (C) the same as (A) but considering the distributions of $\Delta_{\mathrm{z}_i}\mathrm{log}(\psi_*)$. 
	\end{flushleft}
\end{table*}

Finally, we note the existence in each stellar mass bin and in each redshift sub-sample of a population of disk galaxies with observed $\psi_*$ much lower than the median (here we consider the population with $\Delta_{\mathrm{z}_i}\mathrm{log}(\psi_*) <-0.5$). The fractional size of this population increases with stellar mass from $5-10$\% in the low stellar mass bin to $12-17$\% in the highest stellar mass bin, however, is largely constant as a function of redshift. As discussed in \citet{GROOTES2014} and \citetalias{GROOTES2017} this population is dominated by genuine UV faint disk galaxies. The presence of such a low $\psi_*$ population amongst the disk galaxies of the \textsc{FCS} sample, with fractional size increasing with stellar mass, may imply the existence of a secular mechanism, dependent on galaxy (stellar) mass, acting to shut down star formation in disk galaxies.\newline

In summary, we find evolution of the $\psi_*$--$M_*$ relation of disk galaxies over the redshift range $0<z<0.13$, sampled at intervals of $\Delta z \approx 0.03-0.04$, to be consistent with a smooth gradual evolution of the normalization and possibly the slope while the scatter, i.e. the distribution around the median, remains constant. Furthermore, we find the observed evolution to be fully consistent with the extrapolation of the parameterization of the MS presented by \citet{SPEAGLE2014}.\newline

\section{Impact of Environment/DMH mass on the $\psi_*$--$M_*$ relation for central disk galaxies}\label{results_env}
For central galaxies theory predicts the maximum achievable rate of accretion onto the galaxy to be a function of the host DMH mass. However, In \citetalias{GROOTES2017} we have presented evidence that the $\psi_*$--$M_*$ relation for central disk galaxies of groups coincides with that found for their field counterparts. While this result may imply a possible lack of halo mass dependence
 of the $\psi_*$--$M_*$  relation for central disk galaxies, it may also arise from the fact that the DMH masses of field and group central disk galaxies at fixed stellar mass are highly similar. Here, we focus on comparing the  $\psi_*$--$M_*$ relations for our samples of field and group central disk galaxies, the \textsc{FCS} and \textsc{GCS} samples, respectively -- combining this comparison with an investigation of their respective DMH masses to constrain a possible dependence on DMH mass.\newline

As discussed in section~\ref{dataandsamples} and shown in Fig.~\ref{fig_FGCS_spfsm} the range in $M_*$ common to the \textsc{FCS} and \textsc{GCS} samples is limited to $M_*\ge 10^{9.8} M_{\odot}$. However, as is apparent from Fig.~\ref{fig_FGCS_spfsm}, even within this mutual stellar mass range the relative distributions of $M_*$ differ radically, making any average property potentially vulnerable to a bias arising from this dissimilar distribution of stellar mass. We therefore proceed by creating four subsamples -- two from the \textsc{GCS} sample and two from the \textsc{FCS} sample -- with which to perform the brunt of our analysis. Those from the \textsc{GCS} sample are obtained by splitting the sample at its median stellar mass of $\mathrm{log}(M_*/M_{\odot}) = 10.6$,  and we will refer to the low and the high stellar mass sub-samples as the \textsc{LGCS} and \textsc{HGCS} samples, respectively. To construct the two from the \textsc{FCS} sample, analogously referred to as the \textsc{LFCS} and \textsc{HFCS} samples, we begin by selecting all galaxies from the \textsc{FCS} sample with $\mathrm{log}(M_*/M_{\odot}) \ge 9.8$. This sample is then split at $\mathrm{log}(M_*/M_{\odot}) = 10.6$ analogously to the \textsc{GCS} sample. Subsequently, within each stellar mass range, we randomly select galaxies in such a manner as to reproduce the mass distributions of the \textsc{LGCS} and \textsc{HGCS} samples, respectively, while simultaneously maximizing sample size. A detailed description of this process is provided in Appendix~\ref{APPEND_sample}.\newline

\subsection{The $\psi_*$--$M_*$ relation for field and group central spiral galaxies}\label{psistarMstarfgc}
\begin{figure}
	\includegraphics[width=\columnwidth]{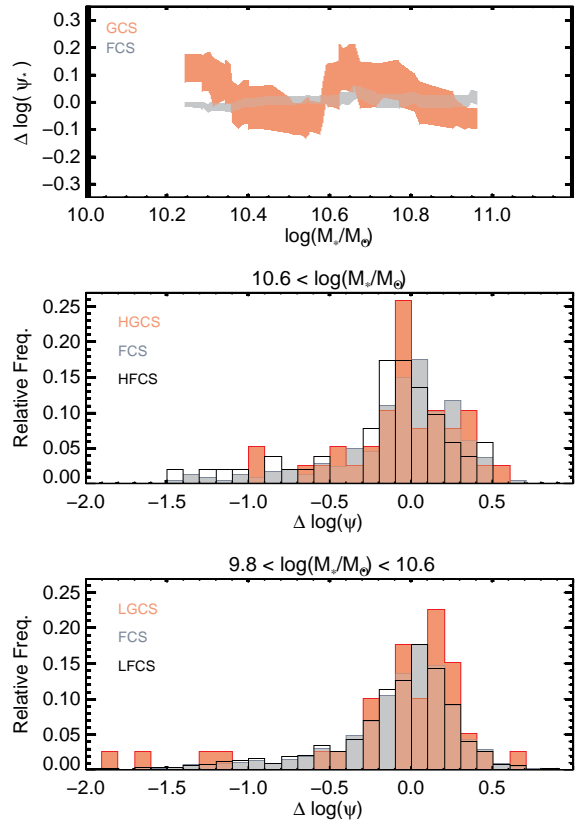} 
	\caption{\textbf{Top}: Median value of $\Delta \mathrm{log} (\psi_*)$ as a function of $M_*$ for the \textsc{GCS} sample in a sliding tophat bin containing 25 galaxies (red shaded area). The width of the shaded region denotes the boot-strapped uncertainty in the median value. The median value of $\Delta \mathrm{log} (\psi_*)$ for the \textsc{FCS} sample in bins covering the range in stellar mass corresponding to that of the sliding tophat applied to the \textsc{GCS} sample is shown as a gray shaded region, with the width again indicating the uncertainty in the median. \newline 
	\textbf{Middle}: Distribution of $\Delta\mathrm{log}(\psi_*)$ for the \textsc{HGCS} sample (filled red histogram), the \textsc{HFCS} sample (open black histogram), and the \textsc{FCS} sample in the mass range $ \mathrm{log}(M_*/M_{\odot}) \ge 10.6$. All three distributions are similar to the degree that the null hypothesis that they are all drawn from the same parent population can not be discarded.\newline
	\textbf{Bottom}: As middle panel but for the stellar mass range $9.8 \le \mathrm{log}(M_*/M_{\odot}) < 10.6$, i.e. the \textsc{LGCS} and \textsc{LFCS} samples. }
	\label{fig_GCSFCS_ssfrsmdlogpsi_p1}	
\end{figure}

We begin our investigation by considering the $\psi_*$--$M_*$ relations of the \textsc{FCS} and \textsc{GCS} samples, as depicted in Fig.~\ref{fig_GCSFCS_ssfrsmdlogpsi_p1} (top panel). We find the median $\Delta \mathrm{log}(\psi_*)$--$M_*$ relations to coincide. This finding is supported by the bottom (middle) panel of  Fig.~\ref{fig_GCSFCS_ssfrsmdlogpsi_p1}, which shows the distributions of $\Delta \mathrm{log}(\psi_*)$ for the \textsc{LGCS} (\textsc{HGCS}) sample, the \textsc{LFCS} (\textsc{HFCS})sample, and the \textsc{FCS} limited to the range $9.8 \le \mathrm{log}(M_*/M_{\odot}) < 10.6$ ($10.6 \le \mathrm{log}(M_*/M_{\odot})$). These all appear to be highly similar, and a 3-sample Anderson-Darling test comparing the distributions finds no grounds to reject the null hypothesis that the distributions are all consistent with having been drawn from the same parent sample ($p=0.4$ and $p\gtrsim0.98$, respectively). In line with this result, we find the scatter in the relations in each range of stellar mass, as characterized by the inter-quartile range of the distribution of $\Delta \mathrm{log}(\psi_*)$, to be consistent between the \textsc{LFCS}, \textsc{LGCS}, \textsc{HFCS}, and \textsc{HGCS} samples, i.e. $0.37\pm0.02\,$dex, $0.3(3)\pm0.10\,$dex, $0.3(4)\pm0.12\,$dex, and $0.3(7)\pm0.10\,$dex, respectively. 

We do note, however, that the $\Delta \mathrm{log}(\psi_*)$ distributions of the \textsc{LGCS} and \textsc{HGCS} samples appear to be shifted slightly towards higher $\psi_*$ compared to the mass-matched \textsc{LFCS} and \textsc{HFCS} samples. so that the $\psi_*$--$M_*$ relation of the \textsc{GCS} sample may be slightly elevated compared to that of the \textsc{FCS} sample.\newline  

Overall, we thus find no robust evidence of a systematic difference in $\psi_*$ at fixed $M_*$ between field and group central disk galaxies.\newline

\subsection{The (average) DMH masses of field and group central spiral galaxies}\label{avDMHmass}
Having found the $\psi_*$--$M_*$ relations of field and group central disk galaxies to coincide, we consider the DMH halos of these galaxies, focussing on their mass. Determining the DMH mass for these sources poses a significant challenge and, in fact, is unfeasible on an individual basis, at least for the field central disks. We therefore estimate average DMH masses for our four sub-samples, using several complementary methods.\newline 

\begin{figure*}
\includegraphics[width=0.7\textwidth, angle=90]{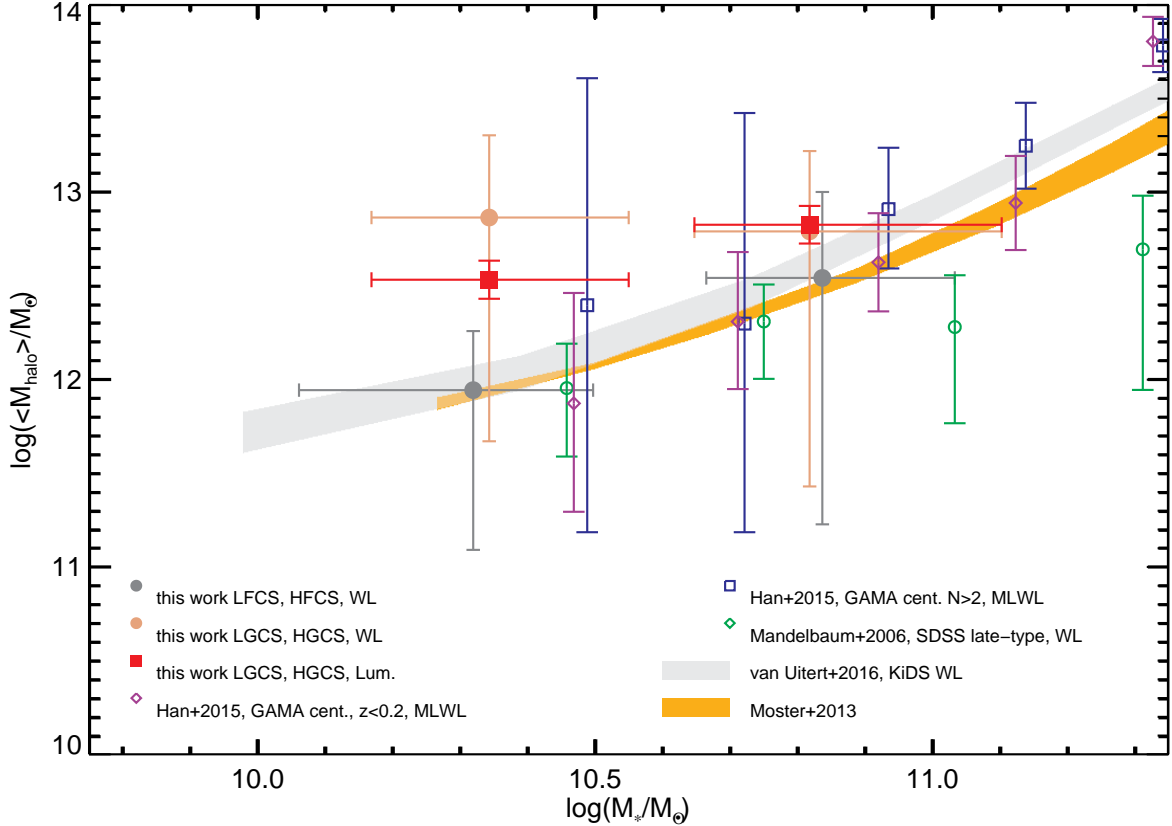}
\caption{Average host DMH mass of the \textsc{LFCS} and \textsc{HFCS} samples (gray filled circles)
as well as of the \textsc{LGCS} and \textsc{HGCS} samples (filled salmon circles) as determined from stacked weak-lensing measurements using the KiDS galaxy-galaxy weak-lensing pipeline as detailed in appendix~\ref{APPEND_WL}. The symbols denote the mode of the posterior PDFs while the error bars show the $68$\% highest probability density interval containing the mode. In stellar mass the symbols denote the median with the error bars indicating the $16$th to $84$th percentile range in stellar mass contributing to the the stack. For the \textsc{LGCS} and \textsc{HGCS} samples the independent group luminosity based median DMH mass estimates are shown as filled red squares with the error bars indicating the uncertainty in the median. For comparison the empirical median $M_{\mathrm{halo}}$--$M_*$ relation derived by \citet{VANUITERT2016} using the full KiDS-GAMA overlap, as well as the abundance matching based $M_*$--$M_{\mathrm{halo}}$ relation of \citet{MOSTER2013} - converted to the form $M_{\mathrm{halo}}$--$M_*$ \citep{HAN2015} - are shown as gray (dash-dotted) and yellow (dashed) outlined regions, with the width indicating the range between the $16$th and $84$th percentile. For further comparison the results of two recent independent comparable weak-lensing analysis are overlaid. The median DMH halo masses found by \citet{HAN2015} using a SDSS imaging based maximum-likelihood weak-lensing analysis in bins of central galaxy stellar mass for a volume-limited (at $z=0.2$) sample of GAMA central galaxies and a flux-limited sample of GAMA group central galaxies ($N\ge3$, i.e. with some similarity to our \textsc{GCS} sample), are shown as open purple diamonds and open blue squares, respectively. Similarly, the median DMH masses of a flux-limited sample of central late-type SDSS galaxies presented by \citet{MANDELBAUM2006b} are shown as open green circles.}
\label{fig_GCSFCS_ssfrsmdlogpsi_p2}
\end{figure*}

An initial expectation value for the average DMH mass of the field central disk galaxies of the \textsc{LFCS} and \textsc{HFCS} samples can be obtained from the stellar-mass -- halo-mass (SMHM) relation. 
 In Fig.~\ref{fig_GCSFCS_ssfrsmdlogpsi_p2} we show the SMHM relation for central galaxies, (i) as derived from KiDS \citep{DEJONG2015,KUIJKEN2015} weak-lensing observations \citep[gray;][]{VANUITERT2016}, and (ii) as expected based on abundance matching applied to the Millennium simulations \citep[orange;][and references therein]{MOSTER2013}\footnote{These have been converted from the usual form $\left<M_*|M_{h}\right>$ to $\left<M_{h}|M_{*}\right>$ using Bayes' theorem for conditional probabilities (\citealt[][]{HAN2015};see also \citealt{COUPON2015}).}. It is, however, initially unclear to what degree the average halo mass of our \textsc{LFCS} and \textsc{HFCS} samples conform to this expectation. 
 
 For each sample, we have therefore constructed a stacked shear profile and corresponding excess surface density profile using the bespoke KiDS galaxy-galaxy weak lensing pipeline and have fit a DMH mass assuming a single NFW halo. Details of this process are provided in Appendix~\ref{APPEND_WL}. As shown in Fig.~\ref{fig_GCSFCS_ssfrsmdlogpsi_p2}, the average DMH masses derived for the \textsc{LFCS} and \textsc{HFCS} sub-samples ($\mathrm{log}(M_{h}/M_{\odot}) = 11.9^{+0.3}_{-0.9}$ and $\mathrm{log}(M_{h}/M_{\odot}) = 12.5^{+0.5}_{-1.3}$, respectively\footnote{The values quoted correspond to the mode and the $68$\% highest probability density(HPD) interval of the posterior distribution.}) are consistent with the expectations. 
 
 We note, however, that the empirical SMHM relation derived by \citet{VANUITERT2016} is based on the flux-limited GAMA sample, while our \textsc{LFCS} and \textsc{HFCS} are volume-limited and morphologically selected, thus introducing the possibility of a systematic bias. For comparison, we therefore show the average DMH mass in bins of stellar mass determined for, (i) a volume limited ($z<0.2$) sub-sample of GAMA central galaxies using maximum-likelihood weak-lensing \citep[purple][]{HAN2015}, and (ii) a flux-limited sample of late-type SDSS central galaxies using stacked weak-lensing \citep[green;][and references therein]{MANDELBAUM2006b}. We find our average DMH mass estimates for both sub-samples to be consistent within the uncertainties with the body of literature. Finally, we note that our limited sample size of 52 (1013) galaxies in the \textsc{HFCS} (\textsc{LFCS}) samples limits the strength of the shear signal and the tightness of our constraint on the average halo mass.
 
Nevertheless, our findings indicate that that the average halo mass of our samples are consistent with expectations and certainly provide a clear upper limit.\newline

As for the field sub-samples, constraining the average DMH mass of the \textsc{LGCS} and \textsc{HGCS} samples using stacked weak-lensing measurements is complicated by the small sample sizes ($40$, and $39$ sources, respectively), and we obtain $\mathrm{log}(M_{h}/M_{\odot}) = 12.8^{+0.4}_{-1.2}$ and $\mathrm{log}(M_{h}/M_{\odot}) = 12.8^{+0.4}_{-1.4}$ for the low and high stellar mass sub-sample, respectively.

However, for these samples we can also make use of the group luminosity to derive an estimate of the mass of the (central) DMH of the group following \citet{HAN2015}. This is done by using the observed group luminosity from the G$^3$C \citep{ROBOTHAM2011b} and Eqs.~22, \& 23 of \citet{ROBOTHAM2011b} to estimate the total group luminosity and subsequently employing the luminosity based halo mass estimator for GAMA presented by \citet{HAN2015}. \newline

Using this estimator we obtain average DMH masses of $\mathrm{log}(M_{\mathrm{DMH}}/M_{\odot}) = 12.5 \pm 0.1$ and $\mathrm{log}(M_{\mathrm{DMH}}/M_{\odot}) = 12.8 \pm 0.1$, for the \textsc{LGCS} and \textsc{HGCS} samples, respectively, as shown in red in  Fig.~\ref{fig_GCSFCS_ssfrsmdlogpsi_p2}. As such, both estimators of the average DMH masses of our group central sub-samples agree within their uncertainties, especially for the \textsc{HGCS} sample. Nevertheless, for the \textsc{LGCS} sample, although the luminosity derived average mass is consistent with the stacked weak-lensing based estimate within the uncertainties of the latter, the average DMH mass preferred by the luminosity based estimate does seem to be slightly lower. Furthermore, the preferred stacked weak-lensing halo masses for both samples are highly similar.\newline 

We note, however, that the average halo mass preferred by the stacked weak-lensing analysis may differ systematically from the median of the luminosity based estimates, due to the underlying $M_h$ distribution and the dependency of the tangential shear signal on DMH mass, in particular for the \textsc{LCS} sample where the shear signal from lower mass halos might be dominated by that from higher mass systems.

Furthermore, we also note that our sample of group central disk galaxies includes a multiplicity based selection, strongly akin to that used by \citet{HAN2015}. In Fig.~\ref{fig_GCSFCS_ssfrsmdlogpsi_p2} we therefore also show the maximum-likelihood weak-lensing based estimates of the halo mass of the multiplicity limited sample of GAMA groups considered by \citet{HAN2015} in bins of central stellar mass (shown in blue). We find our weak-lensing and group luminosity derived halo mass estimates to be consistent with the findings of \citet{HAN2015}, and recover evidence for the dependence of the central SMHM relation on multiplicity at lower central stellar mass observed by these authors.\newline 

Finally, we emphasize, that the halo mass estimator of \citet{HAN2015} is suitable to our consideration of potentially different halo masses at a fixed stellar mass, as it employs group luminosity as a proxy for halo mass, and has been directly calibrated on weak-lensing measurements of GAMA galaxy groups using a selection function closely related to that adopted in our analysis. We refer the inclined reader to \citet{HAN2015} for further details, but also note that, as shown by \citeauthor{HAN2015}, at the low halo mass end any bias in the estimator is likely to underestimate the true halo mass, making our measurements conservative estimates. Furthermore, we note that \citet{VIOLA2015} have presented an updated relation between group luminosity and halo mass using the combined GAMA and KiDS data. These authors find their results, obtained using a more sophisticated halo model, to be fully consistent with those of \citet{HAN2015}. As the halo model we have employed in our analysis corresponds to that of \citet{HAN2015} and our sample is akin to the sample of groups with $N\ge3$ used by \citet{HAN2015} we have made use of the relation presented there, 
but emphasize that our results are robust against the substitution of this relation with that of \citet{VIOLA2015}.\newline

Overall, we thus conclude that the average DMH masses of the low and high stellar mass samples of group central galaxies are $\gtrsim 10^{12.5} M_{\odot}$ and $\gtrsim 10^{12.8} M_{\odot}$, respectively, i.e. comparable to and possibly slightly higher than that of their field central counterparts in the high stellar mass range, and $\gtrsim4 - 8$ times more massive in the low stellar mass range.\newline

In summary, we find that at higher stellar mass both the distributions of $\Delta\mathrm{log}(\psi_*)$ ( and accordingly the median $\psi_*$--$M_*$ relations) and the large scale environment of these galaxies in terms of the mass of their host DMH are similar between the group and field central galaxies.    
In contrast, although in the lower stellar mass range ($9.8 \le \mathrm{log}(M_*/M_{\odot}) < 10.6$) the average DMH mass of group central disks is $\gtrsim 4 - 8$ times greater than that of their field central counterparts, the $\psi_*$--$M_*$ relations and the distributions of $\Delta\mathrm{log}(\psi_*)$ are statistically indistinguishable. Thus, in the low DMH mass range probed by our lower stellar mass group and field central galaxies our results disfavour a halo mass dependence of the $\psi_*$--$M_*$ relation of these galaxies.\newline

\section{Implications for Gas-Fuelling and the Baryon Cycle}\label{IMPGF}
In the previous sections we have demonstrated (i) that the $\psi_*$ - $M_*$ relation for central disk galaxies evolves smoothly with redshift over very short redshift intervals ($\Delta z \sim 0.04$), in line with the expected behaviour of the MS calibrated over much larger redshift intervals (Section~\ref{results_evol}), and (ii) that the $\psi_*$ - $M_*$ relations of group and field central disk galaxies are statistically indistinguishable over their full mutual range in stellar mass, even in the low stellar mass range where the host DMH masses differ by a factor of $\gtrsim 4 - 8$, i.e. we find no evidence for a halo mass dependence of the $\psi_*$ - $M_*$ relation of central disk galaxies, at least in the range of low DMH masses (Section~\ref{results_env}). In the following, we will discuss the implications of these findings in the context of the baryon cycle and the gas-fuelling of these central disk galaxies.\newline
 
 \subsection{The baryon cycle paradigm}\label{IMPGF_BC}
 For (largely isolated) disk central galaxies, such as those in our \textsc{FCS} and \textsc{GCS} samples, whose accretion history is dominated by smooth accretion
, the mass of the ISM and its time dependent evolution can be expressed as
\begin{equation}
\dot{M}_{\mathrm{ISM}} = \dot{M}_{\mathrm{in}} - \dot{M}_{\mathrm{out}} - (1 - \alpha)\Phi_*\,,
\label{eq_ISMevol}
\end{equation}
where $\dot{M}_{\mathrm{in}}$ and $\dot{M}_{\mathrm{out}}$ are the in- and outflow rates of gas from the galaxy, $\Phi_*$ is the SFR and $\alpha$ is the fraction of mass (instantaneously) recycled back to the ISM from high mass stars. As detailed in \citetalias{GROOTES2017}, assuming a volumetric star-formation law (i.e. $\Phi_* = \tilde{\kappa}M_{\mathrm{ISM}}$ \citealp[e.g.][]{KRUMHOLZ2012}), and that the outflow from a galaxy is proportional to its ISM mass \footnote{Under a volumetric star-formation law this is equivalent to the mass-loading formalism.}, Eq.~\ref{eq_ISMevol} can be equivalently formulated in terms of the ISM mass and the SFR as
\begin{eqnarray}
\dot{M}_{\mathrm{ISM}}  & = &\dot{M}_{\mathrm{in}} - \frac{M_{\mathrm{ISM}}}{\tau_{\mathrm{res}}} - \kappa M_{\mathrm{ISM}}\label{eq_MdotMISM}\\
& = & \dot{M}_{\mathrm{in}} - \lambda \Phi_* - (1 - \alpha)\Phi_* \label{eq_MdotPhi}
\end{eqnarray}
where, in Eq.~\ref{eq_MdotMISM} we have cast the constant of proportionality relating the outflow rate to the ISM mass in terms of typical residence time $\tau_{\mathrm{res}}$ for a unit mass of gas in the ISM and have defined $\kappa = (1-\alpha)\tilde{\kappa}$, and in Eq.~\ref{eq_MdotPhi} we have defined the mass loading factor $\lambda = 1/\tau_{\mathrm{res}}\tilde{\kappa}$. We note that $1/\kappa = \tau_{\mathrm{exhaust}}$ corresponds to the gas-exhaustion-by-star-formation timescale in a closed box model. Here, we assume $\tau_{\mathrm{res}}$ and $\tilde{\kappa}$ (and thus $\lambda$) to be determined by galaxy-specific processes, i.e. while they may vary as a function of e.g. galaxy stellar mass, they are constant for all galaxies of a fixed stellar mass.\newline

Accordingly, for the disk galaxies in our samples (read a MS galaxy), provided the inflow is (approximately) stable on timescales longer than the system requires to adjust to perturbations, the SFR of a galaxy is expected to be determined by a self-regulated balance between flows of gas into and out of the ISM and consumption of the ISM via star formation. In this case, the galaxy may be considered to be in a quasi-steady state with a quasi-constant SFR at any given time. Combined with a gradually evolving inflow, this model represents a widely favoured explanation of the observed small scatter in the MS and its gradual evolution with redshift \citep[e.g.][]{KERES2005,DAVE2012,LILLY2013,SAINTONGE2013,MITRA2015} and we will return to this aspect in section~\ref{IMPGF_ZETAZ} below.\newline
 
 In such a self-regulated quasi-steady state $\dot{M}_{\mathrm{ISM}} \approx 0$ in Eq.~\ref{eq_MdotPhi} \footnote{As discussed in \citetalias{GROOTES2017} the actual behaviour of the ISM mass in a quasi steady state will be bracketed by the conditions $\dot{M}_{\mathrm{ISM}}=0$ and $\mu = M_{\mathrm{ISM}}/M_* = \mathrm{const.}$. In the latter case the inflow is higher by a factor of $(1-\alpha) \mu$ with the end point of the inflow being the growing ISM. The functional form, however, is similar in both cases.}, which can then be reformulated as
\begin{eqnarray}
\Phi_* & = & \frac{1}{\lambda + (1 -\alpha)} \dot{M}_{\mathrm{in}} \label{eq_PhiMdot_lambda}\\ 
          & = &  \frac{1}{\frac{1}{\tilde{\kappa} \tau_{\mathrm{res}}} + (1 -\alpha)} \dot{M}_{\mathrm{in}}\label{eq_PhiMdot_taures}\\
          & = & \tilde{\tau} \tilde{\kappa} \dot{M}_{\mathrm{in}} \label{eq_PhiMdot_tildetau}\,,
\end{eqnarray} 
where $\tilde{\tau} = \tau_{\mathrm{res}} \tau_{\mathrm{exhaust}} / \tau_{\mathrm{res}} + \tau_{\mathrm{exhaust}}$. It is then immediately apparent that the star formation is expected to trace the inflow rate in such a case.
\newline

For central galaxies on the MS, the (evolving) inflow is widely surmised to be defined by the (evolving) rate of DM and baryon accretion onto DM halos and an efficiency $\zeta$, the fuelling efficiency, with which accretable gas is delivered to the ISM of the galaxy \citep[e.g.][]{DAVE2012,LILLY2013,BEHROOZI2013a,MITRA2015,RODRIGUEZ-PUEBLA2016}, such that 
\begin{equation}
\dot{M}_{\mathrm{in}} = \zeta \dot{M}_{\mathrm{b,halo}} = \zeta f_b \dot{M}_{\mathrm{halo}}\,,
\label{eq_MdotZeta}
\end{equation}
where $\dot{M}_{\mathrm{halo}}$ is the halo mass accretion rate, $f_b$ is the (cosmological) baryon fraction, and $\dot{M}_{\mathrm{b,halo}}$, accordingly, is the halo baryon accretion rate - generally well approximated in this manner \citep[e.g.][]{VANDEVOORT2011b,BEHROOZI2013,WETZEL2014}.  
As an expectation for the MS, inserting Eq.~\ref{eq_MdotZeta} into Eq.~\ref{eq_PhiMdot_tildetau}, one obtains
\begin{equation}
\Phi_* = \tilde{\tau}\tilde{\kappa}\zeta f_{b}\dot{M}_{\mathrm{halo}} = \tilde{\tau}\tilde{\kappa}\zeta \dot{M}_{\mathrm{b,halo}}\,. 
\label{eq_PhiMdot_zeta}
\end{equation} 
 I.e., assuming $\tilde{\tau}$ and $\tilde{\kappa}$ are fully determined by the galaxy specific processes, the environmental dependence of the MS is encoded in the product $\zeta \dot{M}_{\mathrm{b,halo}}$ of the fuelling efficiency and the halo baryon accretion rate.\newline 

\subsection{Constraints on the DMH mass dependence of gas-fuelling}\label{IMPGF_ZETAMH}
In section~\ref{results_env} we have shown the median $\psi_*$--$M_*$ relations for group and field central galaxies to be statistically indistinguishable over their full mutual range in stellar mass. If the picture of an inflow-driven self-regulated baryon cycle with $\tilde{\kappa}$ and $\tau_{\mathrm{res}}$ determined by galaxy specific processes, as outlined in section~\ref{IMPGF_BC} above 
is to hold for our samples of central disk galaxies, this result entails that the flow of gas into the ISM of the galaxy must be the same.\newline

Expressed in the terms of Eq.~\ref{eq_PhiMdot_zeta} our empirical result thus implies that
\begin{equation}
\zeta \dot{M}_{\mathrm{b,halo}} \approx \mathrm{const.}\,
\label{eq_zetaMhb_const}
\end{equation}
as a function of DMH mass, at least over the halo mass range of $10^{12}  M_{\odot} \lesssim M_{\mathrm{halo}} \lesssim 10^{13} M_{\odot} $ considered here, i.e. our direct empirical results thus require that the fuelling efficiency scales inversely linearly with the halo baryon accretion rate\footnote{Here we have made use of the assumption of the standard paradigm the the inflow rate of gas into the ISM of the galaxy can be expressed as $\dot{M}_{\mathrm{in}} = \zeta \dot{M}_{\mathrm{b,halo}}$. However, even if one only maintains the notion of an inflow driven self-regulated baryon cycle and does not require that the inflow to the ISM of the galaxy can be adequately parameterized using the halo baryon accretion rate, the finding remains relevant in a more general form. Under the broad but reasonable assumption that the inflow to the ISM of the galaxy can be parameterized as 
$\dot{M}_{\mathrm{in}} = \eta_{\mathrm{acc}} \dot{M}_{\mathrm{cool}}$, where $\dot{M}_{\mathrm{cool}}$ represents the rate at which cool accretable gas becomes available at the center of the halo and $\eta_{\mathrm{acc}}$ represents the efficiency with which it is accreted, our result implies  $\eta_{\mathrm{acc}} \dot{M}_{\mathrm{cool}} = \mathrm{const.}$ and thus tightly constrains the joint DMH mass dependence.}. 

As the halo baryon accretion rate is function of the DMH mass, the dependency of $\zeta$ on $M_{\mathrm{halo}}$ (at a given redshift) can then be derived by inserting $\dot{M}_{\mathrm{b,halo}}(M_{\mathrm{halo}},z)$ in Eq.~\ref{eq_zetaMhb_const}. With recent parameterizations of the halo baryon accretion rate \citet{DEKEL2009,MCBRIDE2009,FAKHOURI2010,FAUCHER-GIGUERE2011} favouring an approximately linear or slightly super-linear dependence ($M_{\mathrm{b,halo}} \propto M_{\mathrm{halo}}^{1.06 - 1.15}$), our empirical requirement, representing the first direct constraints on the scaling of $\zeta$ with DMH mass over the range $10^{12} M_{\odot} \lesssim M_{\mathrm{halo}} \lesssim 10^{13} M_{\odot}$,  results in
\begin{equation}
\zeta \approxprop M_{\mathrm{halo}}^{-1.1}\,,
\label{eq_zetaapproxpropmh}
\end{equation}
i.e. a strong halo mass dependence for $\zeta$, where we have adopted the parameterization of $\dot{M}_{\mathrm{b,h}}$ presented by \citet{FAKHOURI2010}.\newline

We note that, qualitatively, our requirement of a strong halo mass dependence of $\zeta$ is consistent with a recent MCMC fit to the global galaxy population of a simple parameterized equilibrium model of the baryon cycle, closely related to the model adopted here, by \citet{MITRA2015}. However, our direct result of $\zeta \approxprop M_{\mathrm{halo}}^{-1.1}$ is slightly steeper than the $\zeta \approxprop M_{\mathrm{halo}}^{-0.75}$ dependence favoured by \citet{MITRA2015} in the same halo  mass range using their indirect approach.

Similarly, our result is also qualitatively consistent with the findings of \citet{BEHROOZI2013a, BEHROOZI2013b}, who find a dependence of the ratio of SFR to halo baryon accretion rate (corresponding to $\tilde{\tau} \tilde{\kappa} \zeta$) $\propto M_{\mathrm{halo}}^{-4/3}$ for halos with $M_{\mathrm{halo}} \gtrsim 10^{11.7}$ using an abundance matching approach combined with halo merger trees and an MCMC driven parameter optimization technique applied to the full galaxy population.

Finally, a comparison of our results with those of the full cosmological hydrodynamical simulations, e.g. those presented by \citet{VANDEVOORT2011}, is interesting. At $z \approx 0$ these authors find the ratio of the rates of accretion of baryons into the ISM and onto the host DMH (i.e comparable to $\zeta$) to initially increase from $M_{\mathrm{halo}} = 10^{11.9} M_{\odot}$ up to $M_{\mathrm{halo}} \gtrsim 10^{12.5}$ before then decreasing $\approxprop M_{\mathrm{halo}}^{-1}$ (their Fig.~2). While the eventual decrease at higher halo masses is consistent with our empirical results, we find no evidence of the initial increase, possibly indicating remaining issues in the model, e.g. the peak being located at too high halo mass.\newline

\subsection{Constraints on the redshift dependence of gas-fuelling}\label{IMPGF_ZETAZ}
Having considered the halo mass dependence of gas-fuelling, i.e. of the fuelling efficiency parameter $\zeta$, in the previous section, we now turn to investigating a second central aspect of the standard baryon cycle paradigm; the evolving normalization of the MS as the result of an inflow-driven self-regulated baryon cycle and an evolving inflow.\newline

Our finding in Section~\ref{results_evol} that the $\psi_*$ - $M_*$ relation for central disk galaxies evolves smoothly with redshift, exactly in line with the evolution predicted for the MS, while new, is, in essence, not a surprising result, as the MS is dominated by disk galaxies in the local Universe \citep[e.g.][]{WUYTS2011}. 

Importantly, however, we have shown the evolution to be gradual and smooth. Given the low redshift ($z \le 0.13$) and small redshift interval ($\Delta z \lesssim0.1$) probed by the $\mathrm{z}_1$, $\mathrm{z}_2$, and $\mathrm{z}_3$ sub-samples of our \textsc{FCS} sub-sample, and with redshift intervals between the samples of only $\Delta z\approx0.04$ (corresponding to only $\approx4\cdot10^{8}\,$yr), we can reasonably expect the efficiency of star formation\footnote{This refers to the efficiency with which ISM is converted to stars in the galaxy, and must be disambiguated from the use of star formation efficiency by e.g. \citet{BEHROOZI2013a}, who use it to refer to the ratio of SFR to the rate at which baryons are accreted onto the host DM halo of a galaxy.} and other galaxy specific processes, i.e. the physics encoded in $\tilde{\kappa}$ and $\tilde{\tau}$ in Eq.~\ref{eq_PhiMdot_zeta}, to be constant over the considered redshift range at a given stellar mass.

Under these assumptions, our empirical finding of a gradually and smoothly evolving SFR (see Table~\ref{tab_norm}) is qualitatively consistent with the picture of an an inflow-driven, self-regulated baryon-cycle determining the SFR of disk central galaxies (MS galaxies), as encapsulated in Eq.~\ref{eq_PhiMdot_tildetau}, and a smoothly evolving inflow.\newline

As outlined in section~\ref{IMPGF_BC}, for central galaxies on the MS, the evolving inflow is widely surmised to be defined by the evolving rate of DM and baryon accretion onto DM halos and the efficiency $\zeta$ \citep[e.g.][]{DAVE2012,LILLY2013,BEHROOZI2013a,MITRA2015,RODRIGUEZ-PUEBLA2016}. Inserting this assumption in Eq.~\ref{eq_PhiMdot_tildetau} results in Eq.~\ref{eq_PhiMdot_zeta} as a description of the baryon cycle and the expectation that the normalization (i.e at fixed $M_*$) of the median $\psi_*$--$M_*$ relation/the MS should evolve as the product $\zeta \dot{M}_{\mathrm{b,halo}}$ over the redshift range probed by our samples. Given this expectation, it is interesting to compare the shift in normalization we find for the $\psi_*$--$M_*$ relation of field central disk galaxies with the expected evolution in the halo  baryon accretion rate, thus empirically constraining the redshift evolution of $\zeta$.\newline

In the following we have adopted the parameterization of the median halo mass accretion rate (and by extension the baryon accretion rate) derived from the Millenium simulation given by Eq.~2 of \citet{FAKHOURI2010}, which takes the form
\begin{equation}
\dot{M}_{halo} \propto M_{\mathrm{halo}}^{1.1} g(z)
\label{eq_fakhouri}
\end{equation}
In addition, for the mutual stellar mass range of the \textsc{FCS} and \textsc{GCS} samples, and in particular for $M_* \approx 10^{10.3}$, i.e the median stellar mass of the \textsc{LFCS} and \textsc{LGCS} samples, we have empirically shown in the previous section~\ref{IMPGF_ZETAMH} that $\zeta \approxprop M_{\mathrm{halo}}^{-1.1}$ at fixed $z$. Accordingly, combining Eqs.~\ref{eq_fakhouri} \& \ref{eq_zetaapproxpropmh}, one finds that, for this stellar mass range, one expects
\begin{equation}
\zeta \dot{M}_{\mathrm{b,halo}} = \zeta M_{\mathrm{halo}}^{1.1}g(z) \approx C M_{\mathrm{halo}}^{\approx0} g(z)\,,
\label{eq_zdepzetaMbh}
\end{equation}
{with the constant of proportionality $C$ potentially depending on redshift at fixed $M_*$. As a result, under the standard paradigm, the evolution with redshift $z$ of the median $\psi_*$--$M_*$ relation at fixed $M_*$ should be largely independent of halo mass, being dominated by the redshift dependence of $g(z)$ and potential redshift dependence of $C$. Furthermore, any residual halo mass dependence may additionally be expected to be negligible, as numerical integration of Eqn.~2 of \citet{FAKHOURI2010} finds the expected mass growth of halos in the mass range covered by our samples ($M_{\mathrm{halo}} \lesssim 10^{13.5}$) over the redshift range considered to be $\lesssim 3\,$\%. Therefore, we choose to make use of the expression for $g(z)$ provided by \citeauthor{FAKHOURI2010} in deriving the redshift dependent change in the halo baryon accretion rate between the $z_1$, $z_2$, and $z_3$ sub-samples of the \textsc{FCS} sample.

Thus, using the expression for $g(z)$ given in \citet{FAKHOURI2010}, we expect the halo baryon accretion rate to decrease by $0.06\,$dex ($0.02\,$dex, $0.04\,$dex)  from $z_3 \rightarrow z_1$ ($z_3 \rightarrow z_2$, $z_2 \rightarrow z_1$), i.e. $\dot{M}_{\mathrm{b,halo}} \approxprop (1+z)^{2.2}$.\newline

We can now compare this expected decrease in the halo baryon accretion rate with the measured decrease in the sSFR over redshift at the fiducial mass $M_* = 10^{10.3} M_{\odot}$ (the median mass of the \textsc{LFCS} sample), as derived from our power law fits of the $\psi_*$--$M_*$ relation for the subsamples of the \textsc{FCS} sample (tabulated in Table~\ref{tab_PLFITS}). We find the measured decrease in sSFR of the field central disk galaxies at this stellar mass to be $0.14\pm0.03\,$dex ($0.04\pm0.01\,$dex, $0.1\pm0.03\,$dex) from $z_3 \rightarrow z_1$ ($z_3 \rightarrow z_2$, $z_2 \rightarrow z_1$). This corresponds to an evolution $\psi_* \propto (1+z)^{\beta}$ with $\beta = 4.9 \pm 0.9$ at the the fixed value $M_* = 10^{10.3} M_{\odot}$\footnote{In section~\ref{IMPGF_ZETAZ} we are using $M_* = 10^{10.3} M_{\odot}$ as the fiducial stellar mass at which the redshift evolution is measured, as this is the median stellar mass of the \textsc{LFCS} and \textsc{LGCS} samples for which we have empirically established a significant discrepancy in host halo mass between the group and field central galaxies combined with a perfectly overlapping $\psi_*$--$M_*$ relation. In section~\ref{results_evol} we have normalized our power law fits at $M_* = 10^{10}M_{\odot}$, the median redshift of the \textsc{FCS} sample. We find the normalization of our fits, corresponding to the evolution of galaxies with $M_* = 10^{10}M_{\odot}$, to decrease by $0.12\pm0.02\,$dex ($0.03\pm0.01\,$dex, $0.09\pm0.02\,$dex) from $z_3 \rightarrow z_1$ ($z_3 \rightarrow z_2$, $z_2 \rightarrow z_1$). This corresponds to an evolution $\psi_* \propto (1+z)^{\beta}$ with $\beta = 4.3\pm0.3$, while a comparable fit to an extrapolation of the \citet{SPEAGLE2014} parameterization of the MS to the same redshift range finds $\psi_* \propto (1+z)^{\beta}$ with $\beta = 4.5\pm0.3$. Thus, our measured evolution of the $\psi_*$--$M_*$ relation is consistent with the data at higher redshifts, although with a slightly flatter slope, which can be understood from Fig.~\ref{fig_FCS_ssfrsm_evol}.}.\newline

Accordingly, in order for the expectation formulated in Eq.~\ref{eq_PhiMdot_zeta} to be consistent with our empirical results, we would require the fuelling efficiency $\zeta$ to decrease as $(1+z)^{2.7}$ for galaxies with stellar masses of $M_* = 10^{10.3} M_{\odot}$, i.e, in the range over overlap between the \textsc{FCS} and the \textsc{LGCS} samples.\newline

Given the novel, direct nature of our results, it is interesting to compare them to other more indirect data driven considerations, as well as to the results of full cosmological hydrodynamical simulations.

Amongst the latter, \citet{VANDEVOORT2011} considered the 
rate of accretion of baryons into the ISM and onto the host DMH of galaxies in their simulations, finding the ratio between the two to decrease between $z\approx2$ and $z\approx0$ (see their Fig.~2). In particular, at a fixed halo mass of $M_{\mathrm{halo}} = 10^{11.9} M_{\odot}$, they find the ratio to decrease by approx $0.5\,$dex between $z=2$ and $z=1$, i.e $\propto (1+z)^(1.04)$ if parameterized as a single power-law over the full redshift range. Given our empirical results, for the model to remain consistent with the data, the evolution of the ratio in the model must vary with redshift. A more detailed comparison of the evolution, however, is not possible due to the lack of (published) simulation results at low(er) redshifts. We urge the simulation community to consider providing more finely sampled (in time/redshift) results at low(er) redshifts, and to divert some further focus to the question of accretion onto galaxies vs. accretion onto halos.

We can also compare our results to those of \citet{BEHROOZI2013a,BEHROOZI2013b}, who find the ratio of SFR to baryon accretion rate (i.e. $\tilde{\tau}\tilde{\kappa}\zeta$) to display a weak time dependence, decreasing with decreasing redshift below $z\approx 0.8$ \citep{BEHROOZI2013a} based on a detailed, empirically constrained combination of halo abundance matching and halo merger trees from numerical simulations.\footnote{Although \citeauthor{BEHROOZI2013a} consider the full galaxy population, as disk galaxies represent the dominant galaxy population for the relevant range in stellar mass \citepalias[e.g.][]{GROOTES2017}, our samples can be considered to be similar.}. In fact, for $M_{\mathrm{halo}} = 10^{11.9}$, i.e the weak-lensing derived DMH mass for the \textsc{LFCS} sample, these authors find the the ratio of SFR to halo baryon accretion rate to decrease by $\approx0.08\,$dex ($\approx0.03\,$dex, $\approx0.05\,$dex) between $z_3 \rightarrow z_1$ ($z_3 \rightarrow z_2$, $z_2 \rightarrow z_1$), i.e. $\propto (1+z)^{2.7}$,} entirely consistent with the evolution of $\zeta$ required by our results\footnote{\citet{BEHROOZI2013a} provide their data in electronic form. We have interpolated the data provided at $z=0.125$, $z=0.095$, and $z=0.05$ using a cubic spline. From the interpolation, for $M_{\mathrm{halo}} = 10^{11.9} M_{\odot}$, we obtain $\mathrm{log}(\mathrm{SFE}) = -0.76$, $\mathrm{log}(\mathrm{SFE}) = -0.71$, and  $\mathrm{log}(\mathrm{SFE}) = -0.68$, respectively, where $\mathrm{SFE}$ is defined as the ratio between star formation rate and halo baryon accretion rate, i.e. is equivalent to $\tilde{\tau}\tilde{\kappa}\zeta$ in Eq.~\ref{eq_PhiMdot_zeta}.}. In this regard it is furthermore noteworthy that, because our sample is comprised of morphologically-selected disks, our results imply that the fall-off of $\zeta$ with redshift for the overall galaxy
population deduced by \citeauthor{BEHROOZI2013a} cannot
be solely driven by an evolution in morphology
from disks to spheroids with redshift.

In conclusion, the cosmological hydrodynamical simulations of \citet{VANDEVOORT2011} are inconsistent with our results with regard to the halo mass dependency of $\zeta$ (see section~\ref{IMPGF_ZETAMH}) and it is unclear whether or to which degree they are consistent with our results with regard to the redshift dependency, possibly hinting that our physical prescriptions for galaxy growth remain incomplete. On the other hand, our data is consistent with the evolution of $\zeta$ with $z$ found in the empirically
constrained analysis of \citet{BEHROOZI2013a}, and qualitatively consistent, in terms of the halo mass dependence, with the empirically constrained analyses of \citet{BEHROOZI2013a} and \citet{MITRA2015}.\newline

Overall, while our measurements can be made to be qualitatively compatible with the predictions of the picture
of star-formation in field central disk galaxies being determined by a gradually evolving inflow-driven self-regulated
baryon cycle, this, requires $\zeta$ to evolve (decline)
as $\propto (1 + z)^{2.7}$ for galaxies with $M_* = 10^{10.3} M_{\odot}$, i.e. that $\zeta$ is
a strong function of redshift. We will return to this, potentially
problematic, finding below. Furthermore, we note that for galaxies with $M_* = 10^{10} M_{\odot}$ we find a preferred evolution of $\zeta \propto (1 +z)^{2.3}$, which, although the results are consistent within their uncertainties, may hint at a dependence of $\zeta$ on $M_*$.

In addition, given our empirical result of the lack of influence of host halo mass on the $\psi_*$--$M_*$ relation of central disk galaxies, maintaining this paradigm requires the fuelling efficiency $\zeta$ to be linearly anti-proportional to the halo baryon accretion rate, and thus to be a strong declining function of $M_{\mathrm{halo}}$. While this is in general agreement with recent indirect derivations based on a consideration of the global galaxy population \citet{BEHROOZI2013a,MITRA2015}, our direct constraints for the morphologically selected samples of disk galaxies considered here favour $\zeta \propto M_{\mathrm{halo}}^{-1.1}$, shallower and steeper than these the previous indirect derivations, respectively. Remarkably, our finding that $\zeta$ is inversely linearly proportional to the halo baryon accretion rate, in essence, implies that there is no detectable effect of the non-linear growth of DM haloes on the growth of stellar mass in disk-dominated galaxies at the centres of these haloes, even at the current epoch, where this effect should be greatest.

The exact physical mechanisms underlying these results remain unclear, and, alternatively, our results may thus also indicate where the current paradigm may represent too much of an oversimplification.\newline 

Regardless of the exact mechanism or mechanisms responsible, however, it is clear, that the similarity in the observed $\psi_*$-$M_*$ relations of group and field central galaxies of comparable stellar mass residing in DMH of discrepant mass places strong constraints on the interplay of galaxy specific and environment dependent processes shaping the $\psi_*$-$M_*$ relation, which in turn are amenable to being used as tests and benchmarks for galaxy evolution simulations. In this vein, the work presented represents the first direct constraint of the fuelling efficiency $\zeta$ in the mass range $10^{12} M_{\odot} \lesssim M_{\mathrm{halo}} \lesssim 10^{13} M_{\odot}$ considered here. 

Similarly, detailed empirical constraints on the evolution of the $\psi_*$-$M_*$ relation represent a powerful means of constraining, and possibly identifying, physical processes on the basis of their evolving contribution to, and impact on, this fundamental scaling relation.\newline

\section{Discussion}\label{discussion}
As detailed above in section~\ref{IMPGF}, while our results can be made compatible with the current paradigm of the baryon ccle of galaxies, they represent strong constraints and pose a challenge for our notion and understanding of the baryon cycle and its constituent processes. In the following we will discuss a number of immediate and broader implications for gas-fuelling, the baryon cycle, including the fuelling efficiency $\zeta$, and the underlying physical processes arising from our results, as well as possible opportunities of further testing the standard paradigm of the baryon cycle which our results entail.\newline

\subsection{Implications for the MS relation}\label{D_MS} 
Firstly, our finding that the median $\psi_*$--$M_*$ relation for field central galaxies evolves gradually and smoothly over a small range in redshift 
, while simultaneously being fully consistent in normalization (and slope) with the empirically parameterized evolution of the MS calibrated over a much larger range in redshift, immediately strongly implies that the evolution of the MS over the longer redshift baselines is likely determined by the same dependencies, including with regard to the larger-scale environment of the galaxy, which drive the evolution observed in our analysis.\newline

\subsection{Co-evolution of $\zeta$ and galaxy properties}\label{D_EVZ_COEV}
Under the initial assumption that the standard paradigm for the baryon-cycle holds (at least for field central disk galaxies),  
the relative constancy of the ratio of SFR to halo baryon accretion rate, i.e. the product $\tilde{\tau}\tilde{\kappa}\zeta$, identified by \citep{BEHROOZI2013a}, implies that $\zeta$ must evolve with redshift in a fashion correlated with the evolution of outflows of gas from galaxies (i.e. anti-correlated with $\tilde{\tau}$), as the mass-loading of these outflows is contained in $\tilde{\tau}$ and varies with redshift. Thus, given the dependencies of $\zeta$ on $M_{\mathrm{halo}}$ and $z$ implied by our results, this requires a tight co-evolution of processes linked to the scales of the galaxy and the host DMH, respectively. \newline

\subsection{Stellar mass and (the redshift dependence of) the fuelling efficiency}\label{D_EVZ_SMDEP} 
In section~\ref{IMPGF_ZETAZ} we have shown that maintaining the gas-fuelling paradigm requires $\zeta$ to be a strong function of $z$, to the extent that at $M_* = 10^{10.3}$ our empirical results require $\zeta \propto (1+z)^{2.7}$, i.e. making $\zeta$ a stronger function of $z$ than of $M_{\mathrm{halo}}$ even, as we have shown in section~\ref{IMPGF_ZETAMH} that $\zeta \approxprop \dot{M}_{\mathrm{b,halo}}^{-1} \propto M_{\mathrm{halo}}^{-1.1}$. 
With $\zeta$ in the formulation of the standard paradigm representing processes regulating the link between the galaxy (its inflows) and the environment on the scale of the DMH and being considered as a function of $M_{\mathrm{halo}}$ and $z$, this finding must be considered problematic, as it is entirely unclear why such processes should display such a strong redshift dependence, while at the same time changes in halo mass and baryon content are expected to be mild over the redshift baselines considered here -- as are possible changes in galaxy specific properties such as star-formation efficiency.

The difficulty inherent in this finding is further compounded by the fact that $\zeta$ would have to evolve significantly with redshift over a wide range in $M_{\mathrm{halo}}$. Under these circumstances, it becomes difficult to accept an evolution such that $\zeta \dot{M}_{\mathrm{b,halo}} = \mathrm{const.}$ at a given $z$, albeit that the constant of proportionality might vary. Alternatively, and also problematically, one would have to assume that the epoch of $z\approx0.1$, i.e the median redshift of our sample, is special in the sense that it just so happens to be the epoch at which we find $\zeta \dot{M}_{\mathrm{b,halo}} = \mathrm{const}$.

Finally, although only with marginal significance, our empirical results on the evolution of the $\psi_*$--$M_*$ relation do indicate an evolving slope of the relation, as was also observed by \citet{SPEAGLE2014}. This finding is further borne out by our direct consideration of the implied redshift evolution of $\zeta$ at different values of $M_*$, which indicates an evolution varying as a function of stellar mass $M_*$. 

Our results, therefore, provide an indication that the fuelling efficinecy $\zeta$ may, in fact, not be adequately described as a function of $M_{\mathrm{halo}}$ and $z$, i.e. $\zeta(M_{\mathrm{halo}},z) \approx M_{\mathrm{halo}}^{-1.1} g(z)$, but instead may also depend on the stellar mass of the galaxy, i.e. $\zeta(M_{\mathrm{halo}},z, M_*) \approx M_{\mathrm{halo}}^{-1.1} \tilde{g}(z, M_*)$. We will return to this question in future work, but note that a functional dependence of this type, while easily preserving the ability to capture the empirical halo mass dependence of $\zeta$, may allow for a shallower redshift dependence.
\newline

\subsection{Influence of the gravitational potential of the host DMH on $\zeta$}\label{D_ZETAMH_GRAV}
Fundamentally, any inflow into the ISM of a galaxy must be sourced from gas sufficiently cool to be accreted, agnostic of its origin. In this context $\zeta \dot{M}_{\mathrm{b,h}}$  simply parameterizes the flow of cool accretable baryons to the the central galaxy as a function of rate with which baryons are accreted onto the DMH.

As discussed in section~\ref{IMPGF_ZETAMH}, in order to maintain the standard baryon cycle paradigm our results require $\zeta \approxprop \dot{M}_{\mathrm{b,halo}}^{-1} \approxprop M_{\mathrm{halo}}^{-1.1}$ resulting in inflow rates to the central disk galaxies being invariant as a function of halo mass. The differing ratios of stellar mass to DMH mass for the group and field central disk galaxies, and the corresponding differences in the gravitational potential wells of the host DMHs, however, have a range of implications for the availability of hot and cool gas in the host DMH halos of the central galaxies.

On the one hand, the rate of accretion of baryons onto the DMH is expected to increase with the mass of the DMH, thus increasing the total available gas mass. Conversely, depending on its mass, the host DMH of a galaxy will be possessed of a hot pressure supported atmosphere and a shock at or near the virial radius heating the infalling IGM.

Theory predicts the threshold halo mass for the existence of a stable shock, and thus a hot atmosphere, to be $\sim10^{11.7}M_{\odot} - 10^{12} M_{\odot}$, depending on redshift and metallicity \citep[e.g.][]{BIRNBOIM2003,KERES2005,KERES2009,VANDEVOORT2011}. Above the threshold mass, an increasing fraction of the baryon mass accreted onto the DMH is expected to be shocked, requiring it to cool prior to being available for accretion into the ISM of the central galaxy, with the cooling efficiency decreasing with DMH mass. Similarly, the fraction of unshocked/unheated gas which can be accreted directly is expected to decline with increasing DMH mass above the threshold halo mass\citep[e.g.][]{KERES2005,KERES2009,VANDEVOORT2011a,NELSON2013}.

A quantitative assessment of the impact of the differing gravitational potentials on the availability of cold accretable gas has recently been provided by \citet{FAUCHER-GIGUERE2011} who analysed a suite of cosmological hydrodynamical simulations and found, cast in the simpler framework of the baryon cylce formulation adopted here, $\zeta_{\mathrm{grav}} \propto M_{\mathrm{halo}}^{-0.25}$ \citep{DAVE2012,MITRA2015} with an onset of this effect above $M_{\mathrm{halo}} \approx 10^{11.75} M_{\odot}$, where $\zeta_{\mathrm{grav}}$ describes the halo mass dependence of the impact of gravitational heating processes on the fuelling efficiency. Thus, our empirical result of $\zeta \approxprop M_{\mathrm{halo}}^{-1}$ implies some additional process that reduces the rate at which gas is accreted onto the central disk galaxy beyond the impact of the gravitational potential of the host DMH considered above, but which nevertheless scales with the DMH mass in a manner resulting in comparable inflows for field and group central disk galaxies at fixed stellar mass.\newline

Before discussing further physical implications of this finding we first consider the impact of the assumptions of our analysis.
In particular,  we have assumed that $\tau_{\mathrm{res}}$ and $\tilde{\kappa}$, i.e., the residence time of a unit mass of gas in the ISM and the efficiency with which stars are formed from the ISM, depend solely on galaxy properties and are independent of the environment of the galaxy. In combination with the expected difference in inflow to the disk central galaxies arising from the impact of the gravitational potential of the halo this gives rise to the tension and the requirement of additional processes highlighted above. However, it may be reasonable to relax this assumption which might suffice to ameliorate the tension arising from the observed similarity of the $\psi_*$--$M_*$ relation of field and group central galaxies without the requirement of further mechanisms - in particular, by allowing for similar SFRs arising from different inflow rates.\newline

For central galaxies of groups, the greater DMH mass will result in a deeper potential well and a higher virial temperature, which may, in turn, result in the medium surrounding the galaxy being more pressurized than that surrounding a field galaxy of comparable stellar mass. However, this would likely act to suppress outflows from the the galaxy leading to an increase, rather than a decrease in $\tau_{\mathrm{res}}$. At fixed star-formation efficiency $\tilde{\kappa}$ this would serve to exacerbate the tension as it would serve to increase the ISM mass and the SFR.\newline
In contrast to the possible impact of the group environment on $\tau_{\mathrm{res}}$, its potential impact on $\tilde{\kappa}$ might indeed serve to lessen the tension implied by our observations. While stellar mass and SFR of star-forming galaxies are tightly correlated, both quantities are also correlated with the metallicity of the galaxy. In fact, a hyperplane in the space spanned by these parameters can be identified which minimizes the scatter from the plane beyond that achievable by the consideration of only two out of three of the parameters as shown by \citet[e.g.][]{MANNUCCI2010,LARA-LOPEZ2010,LARA-LOPEZ2013}. These authors find the SFR at fixed stellar mass to be anti-correlated with the metallicity of the galaxy, possibly as a result of the weaker feedback from lower metallicity O and B stars resulting in a greater efficiency in the conversion of ISM into stars\citep{DIB2011}. Thus, depending on the difference in metallicity between group and field disk centrals, the inflow required to sustain a self-regulated star-formation at a fixed level might be larger for the former compared to the latter.
\newline

For an environment dependence of $\tilde{\kappa}$, driven by a metallicity difference between field and group central disk galaxies, to consistently explain the similarity in SFRs at different estimated inflow rates, however, this difference would have to evolve smoothly and in a highly balanced manner anti-correlated with the difference in host DMH of field and group central galaxies, while simultaneously accounting for any changes in the value of $\tau_{\mathrm{res}}$ as a function of DMH mass. While possible, in principle, this would require a very high degree of finely tuned covariance, for which no mechanism is readily apparent.

Therefore, under the assumption that $\tau_{\mathrm{res}}$ and $\tilde{\kappa}$ are indeed independent of the environment of the galaxy or, as is likely to be the case, do have limited environmental dependencies which, however, are not sufficient to ameliorate the observed discrepancies by themselves, maintaining the paradigm of an inflow-driven self-regulated baryon-cycle, unequivocally requires (an) other mechanism(s) beyond gravitational heating of the IHM to regulate the rate of accretion onto the central galaxy, with this/these mechanism(s) scaling with the mass of the host dark matter halo.\newline

For the global population of galaxies, the result that gravitational heating alone appears to be insufficient to regulate the supply of gas to the central galaxies of massive halos to levels consistent with observations is well known \citep[e.g.][]{CROTON2006}. To accommodate this finding, the current baryon cycle paradigm invokes (a) mechanism(s) for preventive feedback, loosely linked to feedback from AGN although the physics remain unclear, reducing the rate at which gas is accreted onto the central galaxy. Indeed, the recent MCMC fit to the global galaxy population of a simple parameterized equilibrium model of the baryon cycle, closely related to the model adopted here, by \citet{MITRA2015} finds a preferred strong halo mass dependence ($\approxprop M_{\mathrm{halo}}^{-0.5}$) of the unspecific preventive feedback assumed to be loosely linked to AGNs, and a resulting preferred composite fuelling efficiency parameter $\zeta \approxprop M_{\mathrm{halo}}^{-0.75}$ for the halo mass range considered in our analysis. \newline

Here, we have shown that for our morphologically selected samples of disk central galaxies, we, qualitatively consistently with the work of \citet{MITRA2015}, also require a strong dependence of the fuelling efficiency $\zeta$ on the mass of the dark matter halo in order to maintain the baryon cycle paradigm. However, our empirical results, representing the first direct constraints on $\zeta$ in this halo mass range, prefer a steeper dependence $\zeta \propto M_{\mathrm{halo}}^{-1}$ and accordingly, a halo mass dependence of the unspecified feedback mechanism $\propto M_{\mathrm{halo}}^{-0.75}$.

\subsection{The role of AGN feedback}\label{D_PHYSNAT_AGN}
AGN feedback may limit accretion either by (re-)heating the cooling IHM and/or by driving outflows of gas out of the galaxy. While we cannot make detailed statements on the basis of our data, a consideration of our sample construction may be informative for the role of AGN feedback in determining $\zeta$.

In the construction of our samples we have selected against AGN host galaxies using the \citet{KEWLEY2001} BPT criterion. Based on the ratio of emission lines in the optical, the BPT classification efficiently identifies radiatively efficient 'luminous' AGN with efficient accretion. The rejection rate of disk centrals from our samples on the basis of AGN actvity, i.e. the galaxy would have been included save for the classification as an AGN host, is $~10\,$\% for the both \textsc{FCS} and \textsc{GCS} samples.Thus, as a rough estimate, if all of the galaxies in our sample were to contain a central super massive black hole which periodically enters a stage in which it is visible as a BPT classified AGN, the fraction of $10\,$\% AGN over a baseline of ca. 1 Gyr implies a total active phase of $\lesssim10^8\,$yr. Accordingly, this limits the fraction of time over which luminous AGN are injecting energy, and thus limits the relevant AGN feedback modes. 

Our samples thus do not support a \textit{currently active} luminous AGN as the relevant feedback mechanism. However, if the energy input into the IHM during the active phases is sufficient to impact accretion until the next active phase, feedback from luminous AGN may nonetheless be important. In regard to the importance of radiatively inefficient AGN, we can make no statement based on our data, but do note that the star-forming disk galaxies in our samples are not the stereo-typical massive, elliptical galaxies with low gas-to-stellar mass ratios generally associated with 'maintenance mode' feedback from radiatively inefficient AGN.\newline

\subsection{Gas-fuelling: inferred properties and possible processes}
Of foremost importance amongst our results, the smooth evolution over short timescales observed for the $\psi_*$--$M_*$ relation, and the fact that it so closely traces, and is well parametrized by, the evolution of the halo accretion rate, implies that the IHM of the galaxy's host DMH appears to serve as a buffer for star formation only in a very limited capacity. This finding has significant implications for the physical mechanism(s) underlying $\zeta$.

Importantly, in the adopted formulation, $\zeta$ embodies the efficiency with which \text{all} baryons are accreted into the ISM of the central galaxy, i.e. both those newly accreted onto the DMH from the IGM and thus intrinsically linked to the halo accretion rate, as well as those already present in the DMH/being recycled. To first order, these two 'sources' can be mapped to the cold and hot accretion modes, respectively. Conceptually, at any given time, (a) some fraction of in-falling baryons cools efficiently/remains cold and is rapidly accreted onto the galaxy (i.e. cold mode), and (b) some fraction of the baryons contained in the IHM of the galaxy cools and is subsequently accreted (i.e. hot mode), thus constituting the total accretion.

As our \textsc{FCS} sample extends to halo masses $M_{\mathrm{halo}} \gtrsim 10^{12.8}M_{\odot}$, i.e. from halo mass for which cold mode accretion is expected to be dominant to such systems for which a significant contribution to the fuel for recent star formation from gas that has been heated and has subsequently cooled is expected \citep[e.g.][]{KERES2005,KERES2009,VANDEVOORT2011b}, our empirical results entail that the mechanism(s) encoded in $\zeta$ link(s) the accretion efficiency of baryons into the ISM via \textit{both} modes to the instantaneous halo baryon accretion rate. \newline

For low mass DMHs, which are expected to lack a virial shock and a virialized atmosphere, the dominant cold accretion mode ostensibly meets the formulated requirements, although even here, it is not clear that the current SFR of the galaxy should trace the instantaneous halo baryon accretion, i.e. the accretion of baryons at a distance of $\sim r_{\mathrm{200}}$. However, the uncertainties in the SFR and accretion rate determinations, as well as the ensemble averaging in the determination of the median values, may blur out the anticipated lag.

For more massive DMHs, on the other hand, an appropriate mechanism is less apparent. One possibility, is that the dominance of cold mode accretion extends to larger halo masses in the local universe than suggested by the results of \citet{VANDEVOORT2011b,NELSON2013}. Indeed, using the Omega25 suite of cosmological simulations \citet{WETZEL2014} find that cooling is highly efficient for halos with $M_{\mathrm{halo}}=10^{11}-10^{12} M_{\odot}$ and $M_{\mathrm{halo}}=10^{12}-10^{13} M_{\odot}$ and that the rate of baryon accretion at all radii from the virial radius $r_{\mathrm{vir} }$ down to $\sim40\,$kpc traces that at $r_{\mathrm{vir}}$. However, these authors only considered very limited feedback in their simulations.

A possible additional cooling mechanism enabling such a scenario might be the additional efficient cooling of the IHM by dust injected into it by feedback from the galaxy. As \citet{MONTIER2004} have shown that dust cooling can exceed gas-phase cooling in the IHM for dust-to-gas ratios $\ge 10^{-4}$ by mass (ca. 1\% of the ISM value), this may prove feasible. In addition, a mechanism of this type, could also potentially give rise to a two phase IHM, thus meeting the requirement identified in \citetalias{GROOTES2017} in the context of the SFRs of satellite disk galaxies.\newline 

Alternatively, one might speculate whether the expected filamentary accretion of cold IGM onto the halo might provide the required mechanism even in higher mass halos. Numerical experiments find the filamentary inflows of cold gas to penetrate the hot gas halo to a degree varying with the halo mass before breaking up \citep[e.g.][]{VANDEVOORT2011b,NELSON2013}. If hot halo gas could  cool, condensate, and be entrained in the wakes of the cold clumps formed by the disrupting penetrating filament, e.g. as suggested for the galactic fountain model \citep{FRATERNALI2008,MARINACCI2010,ARMILLOTTA2016}, this would provide a mechanism linking the accretion of cold and hot gas. Again, this process could be supported and enhanced by outflows from the galaxy, enriching the IHM with metals, and in particular dust, possibly reducing the temperature of the hot halo and favouring the survival of cold clumps and the condensation of hot gas. A change in the halo accretion rate would then lead to a change in the effective number of seed clouds, coupling the accretion from the hot IHM component to that of the cold as required. Similarly, increasing the DMH mass would reduce the degree to which the cold streams might penetrate, regulating the prevalence of seed clouds and allowing for a natural halo mass dependence. 

It is, however, unclear whether the efficiency of entrainment and condensation, given the temperatures and densities of the hot halo gas would be sufficient to make this mechanism viable. If this scenario were valid, though, one would expect the number of cold dense gas clumps in the CGM to scale as the product of the halo mass and the ratio of the SFR to the halo mass accretion rate. Ongoing and future studies of the CGM properties, could thus, in principle, provide a further independent test of this speculative mechanism.\newline

\subsection{Implications for rotational support of gas in the IHM}\label{D_rot}
A further important aspect of the accretion efficiency that also remains puzzling, but is strongly constrained by our results, is the contribution/importance of angular momentum. For example, \citet{WETZEL2014} find that in their simulations, at radii below $\sim40\,$kpc, i.e still much larger than the radial extent of the galaxy, the gas experiences significant rotational support. Nevertheless, our results suggest that the final transition of halo baryons through this phase must occur on timescales which are short compared to that on which the halo baryon accretion rate changes in order to maintain the observed correlation between SFR and halo accretion rate.\newline 

\subsection{Beyond merger quenching}
Penultimately, it remains important to note that our result of a strongly negative halo mass dependence of the fuelling efficiency $\zeta$ for samples of disk dominated galaxies clearly demonstrates that mechanisms unrelated to the morphological transformation of galaxies via mergers can play a central role in regulating and potentially eventually quenching star-formation in galaxies. Further support for this result is provided by the population of higher stellar mass quenched disk galaxies in the \textsc{FCS} sample as discussed in \citetalias{GROOTES2017} and section~\ref{results_evol}. \newline

\subsection{Testing the standard paradigm}\label{D_TESTZ}
Finally, we highlight a possible test of the standard paradigm based on the methodology and results presented in this work. 

In the context of the standard paradigm our empirical result that the $\psi_*$--$M_*$ relations of group and field central disk galaxies coincide, while the host DMH masses are discrepant, requires $\zeta \dot{M}_{\mathrm{b,halo}} \approx \mathrm{const.}$. For a fuelling efficiency $\zeta(M_{\mathrm{halo},z})$, if the redshift range of our sample is not unique, in the sense that a similar result is obtained at a different redshift, this poses a strong constraint on the possible differential redshift evolution of $\zeta$ as a function of $M_{\mathrm{halo}}$, as e.g. $\zeta \dot{M}_{\mathrm{b,halo}} = \mathrm{const.}$ would require almost no differential evolution given the near linearity of $\dot{M}_{\mathrm{b,halo}}$ as a function of $M_{\mathrm{halo}}$. If however, the observed evolution of the $\psi_*$--$M_*$ relation requires a differential evolution this would unequivocally entail further dependencies of $\zeta$ (provided $\tilde{\kappa} \tilde{\tau}$ can be assumed to be constant).
Upcoming surveys such as DEVILS\footnote{PI L.J.M. Davies; https://devilsurvey.org} and WAVES \citep{DRIVER2016WAVES} will provide the data required to determine the $\psi_*$--$M_*$ relations of group and field central disk galaxies at higher redshift, thereby facilitating this test.\newline 

A further test will also be enabled by, e.g. WAVES \citep{DRIVER2016WAVES} which will significantly increase the volume over which studies of this type are possible in the local Universe. This provides the unique opportunity of also probing to lower and higher DMH masses to fully cover the range over which a transition between fuelling modes is expected.\newline

\section{summary \& conclusions}\label{summary}
We have presented a detailed investigation of the $\psi_*$-$M_*$ relation for central disk galaxies using purely morphologically selected samples drawn from the GAMA survey, focussing, in particular, on the evolution of the relation over short redshift baselines in the local Universe and on its dependence on the mass of the host DMH of the central galaxy.
In determining DMH masses we have made use of the high resolution imaging data provided by the KiDS survey and the bespoke KiDS galaxy-galaxy weak-lensing pipeline. We present our results in detail as an empirical reference for current and future theoretical and numerical work aimed at understanding the baryon-cycle of these objects. Our investigation has found that:

\begin{enumerate}

\item{The median $\psi_*$--$M_*$ relation of field central disk galaxies is consistent in both normalization and slope with the extrapolation of the empirical parameterization of the main sequence of star forming galaxies presented by \citet{SPEAGLE2014} at the median redshift of the sample.}

\item{The redshift evolution of the $\psi_*$--$M_*$ in the local universe, even over redshift baselines of $\Delta z \approx 0.04$, is characterized by gradual, smooth evolution of the normalization (and possibly the slope) at constant scatter. The observed evolution is entirely consistent (both in normalization and slope) with that expected for the main sequence. }

\item{The median $\psi_*$--$M_*$ relations for field and group central disk galaxies coincide over the full mutual rage in stellar mass and are statistically indistinguishable, with no significant evidence of a an offset between the two. This is mirrored in the distributions of observed SFRs relative to the median SFR at fixed stellar mass for field and group central galaxies split into two ranges of stellar mass.}

\item{By contrast, the average host DMH masses of field and group central disk galaxies differ by $\gtrsim0.6\,$dex ($M_{\mathrm{halo}} = 10^{11.9} M_{\odot}$ and $M_{\mathrm{halo}} = 10^{12.5} - 10^{12.8} M_{\odot}$, for field and group centrals, respectively)  in the low stellar mass range ($10^{9.8} M_{\odot} \le M_* < 10^{10.6} M_{\odot}$, while in the high stellar mass range $M_* \ge 10^{10.6} M_{\odot}$ the average DMH masses are much more comparable ($M_{\mathrm{halo}} = 10^{12.5}$ and $M_{\mathrm{halo}}=10^{12.8} M_{\odot}$, respectively), and in fact compatible with no significant difference).}

\end{enumerate}

Making use of the methods developed in \citet{GROOTES2017} and simple analytical models of the baryon-cycle of galaxies we demonstrate that:

\begin{enumerate}
\setcounter{enumi}{4}
\item{The observed $\psi_*$--$M_*$ relation for central disk galaxies (both field and group centrals) over the full redshift range of our sample ($z\le 0.13$) can be made compatible with the picture of a supply-driven self-regulated baryon-cycle determining the SFR of these galaxies,  including, potentially, the inflow rate of baryons into the ISM being determined by the product of the halo baryon accretion rate $\dot{M}_{\mathrm{b,halo}}$ and the gas-fuelling efficiency $\zeta$.}

\item{However, attaining this compatibility not only requires the the gas-fuelling efficiency $\zeta$ to be a strong function of redshift, i.e. $\zeta \propto (1+z)^{2.7}$ at $M_* = 10^{10.3}M_{\odot}$
 for $z=0-0.13$, but also requires $\zeta \dot{M}_{\mathrm{b,halo}} \approx \mathrm{const.}$ as a function of dark matter halo mass, i.e., with $\dot{M}_{\mathrm{b,halo}} \approxprop M_{\mathrm{halo}}^{1.1}$, $\zeta \approxprop M_{\mathrm{halo}}^{-1.1}$. This entails that the gas-fuelling rate of central galaxies is largely independent of the host DMH mass, thus supplying strong constraints on the possible underlying mechanisms.}

\end{enumerate}

Based on these results, we derive a number of further implications and constraints for the physical mechanism(s) underlying the gas-fuelling efficiency $\zeta$.
In particular, we find that

\begin{enumerate}
\setcounter{enumi}{6}
\item{Regardless of the underlying mechanism, the near perfect recovery of the extrapolated MS by the $\psi_*$--$M_*$ of field central disk galaxies most strongly implies that the smooth evolution observed for this relation between $z\approx0.2 -4$ is also determined by the same processes on the scale of the surrounding environment constrained by our local Universe sample.}

\item{A joint consideration of the ratio of SFR to halo baryon accretion rate, the $\psi_*$--$M_*$ relation, and the dark matter halo mass for group and field central galaxies over a redshift baseline extending to intermediate redshifts ($z\gtrsim0.4$) provides a sensitive test/constraint of the standard paradigm for the baryon cycle of star-forming galaxies.}

\item{Under the standard paradigm, the gas fuelling efficiency $\zeta$ must evolve with redshift in a manner correlated with the evolution of the strength of the outflows from the galaxy, requiring a link between galaxy scale and DM halo scale processes. The physical mechanisms underlying $\zeta$, both with regard to the required link to between galaxy and DM halo scale processes, as well as driving the $\propto (1+z)^{2.7}$ redshift dependency remain unclear. We note, however, that a functional dependency of $\zeta$ on $M_*$, $z$, and $M_{\mathrm{halo}}$ (rather than just on $z$ and $M_{\mathrm{halo}}$) may allow for a shallower redshift dependence of $\zeta$.}

\item{The reservoir of gas constituted by the intra-halo medium (IHM) only acts as a buffer for star-formation in a very limited, largely negligible manner as the SFR so closely traces the instantaneous halo baryon accretion rate. In essence this implies that the combined system of IHM and ISM is in an equilibrium state, with a timescale on which the equilibrium adjusts that is short compared to the rate of change of the cosmological inflow to the halo.}

\item{Either cold mode accretion remains the dominant process in delivering gas for star for star formation to the ISM to higher DMH masses than currently thought ($\gtrsim 10^{12.8} M_{\odot}$), or the processes determining $\zeta$ link hot and cold mode accretion to the instantaneous halo baryon accretion rate equally. In both contexts we suggest that cooling of the IHM by dust injected via stellar feedback may serve to increase the cooling rate and decrease the temperature of the IHM leading to and aiding in the formation of a multi-phase (at least two) ISM.}

\end{enumerate} 

Overall, our analysis finds the SFR of central disk galaxies, and in particular field central disk galaxies, can be made to be compatible with the standard paradigm of the baryon cycle of these galaxies which dominate the MS relation being determined by a inflow-driven self-regulated baryon-cycle, with the inflow depending on the evolving halo accretion rate and gas-fuelling efficiency, albeit with a significant degree of tension. In particular, we find that, in order to be compatible, our results very strongly limit the possible influence of halo mass on the SFR of central disk galaxies, requiring the efficiency of gas-fuelling to scale anti-proportionally to the  
halo baryon accretion rate. Furthermore our results would require a strong redshift of the fuelling efficiency. The underlying mechanisms of such dependencies, however, remain unclear. 

The results of our analysis clearly indicate that a wide range of processes determining the cycle of baryons through galaxies remain poorly understood. Nevertheless, in this work, we have provided a number of detailed, both qualitative and quantitative, constraints for current and future theoretical and numerical work addressing the question.\newline

\section*{Acknowledgements}
We thank the referee Romeel Dav\'e for his insightful comments which helped us improve the paper.
MWG would like to thank Janine Pforr for useful discussions.  
CCP acknowledges support from the Leverhulme Trust Research Project Grant RPG-2013-418 and from a previous grant from the UK Science and Technology Facilities Council (STFC; grant ST/J001341/1).
GAMA is a joint European-Australasian project based around a spectroscopic campaign using the Anglo-Australian Telescope. The GAMA input catalogue is based on data taken from the Sloan Digital Sky Survey and the UKIRT Infrared Deep Sky Survey. Complementary imaging of the GAMA regions is being obtained by a number of independent survey programs including GALEX MIS, VST KiDS, VISTA VIKING, WISE, Herschel-ATLAS, GMRT, and ASKAP providing UV to radio coverage. GAMA is funded by the STFC (UK) , the ARC (Australia), the AAO, and the participating institutions. The GAMA website is: http://www.gama-survey.org. \newline
GALEX (Galaxy Evolution Explorer) is a NASA Small Explorer, launched in April 2003. We gratefully acknowledge NASA's support for construction, operation, and science analysis for the GALEX mission, developed in cooperation with the Centre National d'Etudes Spatiales (CNES) of France and the Korean Ministry of Science and Technology.\newline
We thank the KiDS Consortium for making public the weak lensing catalogues that were used in this analysis at this website http://kids.strw.leidenuniv.nl/DR3/lensing.php  
This research is is based (in part) on data products from observations
made with ESO Telescopes at the La Silla Paranal
Observatory under programme IDs 177.A-3016, 177.A-3017
and 177.A-3018.




\bibliographystyle{mnras}
\bibliography{grootes_SPGasfuelling_field_MNRAS}



\appendix

\section{Selection of the \textsc{LFCS} and \textsc{HFCS} samples}\label{APPEND_sample}
For our comparative investigation of the impact of DMH mass on the $\psi_*$--$M_*$ relation of central disk galaxies and their corresponding distributions of $\Delta\mathrm{log}(\psi_*)$ we require samples of field and group central disk galaxies constructed in such a manner as to alleviate the impact of the very different stellar mass functions of these objects, i.e. mass-matched samples. However, the steeply declining mass function of the \textsc{FCS} sample simultaneously limits the maximum attainable sample of unique galaxies. In constructing our samples we have therefore adopted the approach outlined in the following.\newline

We begin by splitting the sample of group central disk galaxies at its median stellar mass, creating a low mass (\textsc{LGCS}) and a high mass (\textsc{HGCS}) sub-sample. For each of these samples we compute the relative frequency of sources in bins of $0.2\,$dex width in $M_*$. From the \textsc{FCS} sample we then select all galaxies which reside in the stellar mass range of the \textsc{LGCS} and \textsc{HGCS} samples, respectively. As for the group central samples, we determine the relative frequency in bins of $0.2\,$dex width in $M_*$ and identify the least populous bin, which we refer to as the normalization bin. From this bin we select all galaxies, and then proceed to select samples of galaxies from all other bins by randomly drawing (without replacing) a number of galaxies corresponding to the relative frequency of the \textsc{LGCS} or \textsc{HGCS} sample for that mass bin, multiplied by the number of galaxies in the normalization bin. This procedure results in the mass matched \textsc{LFCS} and \textsc{HFCS} sub-samples which we make use of in our comparison. We note that in terms of sample size, the \textsc{LFCS} and \textsc{HFCS}  sub-samples represent the maximum size mass-matched samples of unique sources that can be constructed from the \textsc{FCS} sample.\newline 

\section{Stacked weak-lensing DMH mass estimates using the KiDS galaxy-galaxy lensing pipeline}\label{APPEND_WL} 
The GAMA survey area used in the analysis presented overlaps with the KiDS \citep{DEJONG2015,KUIJKEN2015,DEJONG2017} ESO public survey using the VST, thus providing imaging of sufficient quality to allow for a stacked weak lensing analysis of our samples of field and group central disk galaxies in both the low and high stellar mass ranges. This makes use of cosmic shear measurements from KiDS \citep{KUIJKEN2015,HILDEBRANDT2017,FENECH_CONTI2017}. The KiDS data are processed by THELI \citep{ERBEN2013} and Astro-WISE \citep{BEGEMAN2013,DEJONG2015}. Shears are measured using lensfit \citep{MILLER2013}, and photometric redshifts are obtained from PSF-matched photometry and calibrated using external overlapping spectroscopic surveys (see \citealp{HILDEBRANDT2016}).

In our analysis we have made use of the bespoke KiDS galaxy-galaxy weak lensing pipeline which encompasses the creation of an azimuthally averaged excess surface density (ESD) profile. We refer the reader to the dedicated works of the KiDS collaboration on weak-lensing \citep[e.g.][]{VIOLA2015,SIFON2015,BROUWER2016,VANUITERT2016,DVORNIK2017} for a detailed description of the process. In a second step, the pipeline enables a halo model to be fit to the stacked ESD profile making use of the affine invariant MCMC ensemble sampler proposed by \citet{GOODMANWEARE2010} and implemented in Python by \citet{FOREMAN-MACKEY2013}. Given our relatively small sample sizes and our selection of central galaxies we adopt a single NFW profile \citep{NFW} as a halo model, using the mass-concentration relation of \citet{DUFFY2008}, i.e. with the DMH mass as a free parameter. Here we define the mass of the halo as $M_{200}$, i.e. the mass contained within the radius $r_{200}$, for which the average density within the radius is 200 times the mean background density at the median redshift of our sample $z=0.1$. For the fit we use $150$ walkers with a chain length of $50000$ steps, discarding the first 2000 steps as burn-in. For the DMH mass we choose a flat prior in log-space with $10 < log(M_{200}/M_{\odot}) < 15$. We report our results as the mode of the distribution, and the boundaries of the 68th percent highest probability density interval (surrounding the mode).

In modelling the mass distribution with single NFW profile we ignore any contribution from the central stellar component, i.e. the visible galaxy. However, this is, at least partially accounted for by the lower limit on our prior of $log(M_{200}/M_{\odot}) = 10$. Furthermore, this assumption is generally found to be reasonable, at least at the level of $\lesssim0.1\,$dex in terms of derived mass, and likely better \citep[e.g.][]{MANDELBAUM2006b,HAN2015,VANUITERT2016}, in particular given the small size of our stacks and the according inherent low S/N.\newline

\bsp	
\label{lastpage}
\end{document}